\documentclass[twocolumn,secnumarabic,amssymb, nobibnotes, aps, prd]{revtex4-2}
\usepackage{amsmath}
\usepackage{amsfonts}
\usepackage{amssymb}
\usepackage{graphics}
\usepackage{graphicx}
\providecommand{\U}[1]{\protect\rule{.1in}{.1in}}

\begin{document}
	
\title{Nondegenerate Josephson Mixers with Enhanced Bandwidth and Saturation Power for Quantum Signal Amplification and Transduction}
\author{Baleegh Abdo}
\author{Dongbing Shao}
\author{Shayne Cairns}
\author{Jae-woong Nah} 
\author{Oblesh Jinka}
\author{Srikanth Srinivasan}
\author{Thomas McConkey}
\author{Vincent Arena}
\author{Corrado Mancini}
\affiliation{IBM Quantum, IBM Research Center, Yorktown Heights, New York 10598, USA.}
\date{\today}

\begin{abstract}
Nondegenerate Josephson mixers (JMs), formed by coupling two different transmission-line resonators to Josephson ring modulators (JRMs), are vital and versatile devices capable of processing microwave signals at the quantum limit. Owing to the lossless nondegenerate three-wave mixing process enabled by the JRM, JMs can perform phase preserving amplification of quantum signals, generate two-mode squeezed states, and perform noiseless frequency conversion. However, due to their limited bandwidth and saturation power, such resonator-based JMs are generally unable to simultaneously process frequency-multiplexed signals required in large quantum processors. To overcome this longstanding dual challenge, we redesign the JRM parameters by optimizing its inductances to suppress higher order mixing products and engineer its electromagnetic environment by incorporating lumped-element coupled-mode networks between the JRM and the two distinct ports of the JM. By implementing these strategies, we measure for JMs realized with four coupled modes per port, operated in amplification (conversion), bandwidths of about $400$ MHz ($700$ MHz) with power reflections above $10$ dB (below $-10$ dB) and saturation powers of about $-110$ dBm at $15$ dB ($-91$ dBm at $-26$ dB). Similarly, we demonstrate for a low external quality factor resonant-mode JM operated in conversion, a maximum bandwidth of about $670$ MHz with power reflections below $-10$ dB and a maximum saturation power of about $-86$ dBm at $-17$ dB. Such nondegenerate JMs with enhanced bandwidths and saturation powers could serve in a variety of frequency-multiplexed settings ranging from high fidelity qubit readout and unidirectional routing of quantum signals to generation of remote entanglement with continuous variables.     
\end{abstract}

\maketitle
\newpage

\section{Introduction}

Nonlinear Josephson-based devices capable of amplifying quantum microwave signals with minimum added noise set by quantum mechanics \cite{YukeJPA,Caves,NoiseReview}, play a pivotal role in the readout chains of superconducting quantum processors \cite{SCquantumOutlook,BuildingLogicalQubits,DispersiveReadout}. When used as first-stage amplifiers, employing three-wave or four-wave mixing processes, they remarkably enable real-time, high-fidelity, single-shot, QND readout of qubit states \cite{VijayQuantumJump,VijayRabiOsc,QuantBAScience,CavityBabySitting}. Moreover, since such nonlinear mixing devices can operate, under certain conditions, as noiseless frequency converters \cite{JPCconv,SNAILcoupler}, they could potentially play other crucial roles in large quantum processors. Such as transducing quantum states between different frequencies \cite{QuantumNode,ConvNIST,SQUIDconvYale} or forming on-chip nonreciprocal devices that replace bulky off-chip magnetic isolators commonly utilized in the protection of qubits against noise coming down the output chain \cite{CircReview,TWPATLiso,CircFCLehnert,JPCDAcirc,Naamancirc,ActiveIso,SMOC}.

In recent years, two prominent and distinct platforms of low-loss and low-noise Josephson parametric devices have emerged. The first is resonator based. It embeds a Josephson nonlinear medium, such as dc-SQUIDs \cite{JPAHat}, SNAILs \cite{SNAILresAmp} or Josephson ring modulators (JRMs) \cite{JPCnature,microstripJPC} into microwave resonators. JRM-based parametric devices are particularly advantageous owing to their nondegnerate nature, i.e., having two spectrally different modes with two separate physical ports \cite{NatPhysJPC,JPCreview} and their three-wave mixing operation in which the pump tone frequency lies out of the device band and can be fed to it via a separate port \cite{hybridLessJPC}. The second is transmission-line based. It forms a semi-continuous nonlinear transmission line made of an array of hundreds to thousands of Josephson elements, such as Josephson junctions \cite{TWPAScience,FloquetTWPA,LowLossTWPA,HighSatTWPA}, SNAILs \cite{ObsTwoModSqTWPA,BroadbandCVKerrJMetamaterial}, and rf-SQUIDs \cite{rfSQUIDTWPAthy,rfSQUIDTWPAExp}, that are capacitively coupled to ground and periodically interleaved in some designs with linear resonant circuits, primarily engineered to enable phase matching between the various traveling waves taking part in the mixing process \cite{ResonantMatchingTWPA}.   

Owing to these basic differences in the engineered electromagnetic environment, the two platforms of Josephson parametric devices exhibit different figures, in particular with respect to added noise, insertion loss, maximum gain or transmission, bandwidth, and saturation power. While the resonator-based devices generally exhibit better performance with regard to the first three metrics, they tend to have very narrow bandwidths and low saturation powers, which preclude them from being used in hardware-efficient, frequency-multiplexed schemes required in large quantum processors. In contrast, the transmission-line devices, due to their distributed nature, generally achieve broad bandwidths spanning a few gigahertz and large saturation powers exceeding $-110$ dBm, which make them indispensable in large quantum systems \cite{TWPAScience,FloquetTWPA,LowLossTWPA}. 

However, despite the practical utility and reliable performance of traveling-wave parametric devices (TWPAs), they possess a few disadvantages when deployed for qubit readout. While their broad bandwidth of $2-5$ GHz is generally considered a major advantage, it is also a liability since it requires for its proper operation well-matched $50$ Ohm environment at its input and output ports over the same bandwidth and amplifies more noise than necessary over its wide bandwidth, including the unwanted qubit frequency band, while most qubit readout signals reside in a much narrower band of about $1$ GHz or less \cite{MultiplexReadoutwithJPA}. As a result, it is imperative to pair TWPAs with broadband isolators to provide sufficient broadband impedance matching and qubit protection against reflected amplified signals or noise \cite{TWPAScience}. Another notable disadvantage of typical TWPAs is the unintended generation and phase-matching of higher-order mixing products \cite{ObrienPhDThesis}, which result in reduction of the qubit measurement efficiency and could cause instabilities if the device ports are not well matched out of band as well.   

One promising scheme to overcome some of these drawbacks applies impedance-matching techniques to resonator-based devices, which ideally boost their bandwidths without compromising their core strengths \cite{SynParamCouplNet}. However, all experimental realizations to date 
achieve bandwidth enhancement in amplification and conversion via impedance matching of grounded Josephson elements, such as dc-SQUIDs and interleaved arrays of rf-SQUIDs. Furthermore, most experimental works apply impedance-matching to doubly degenerate parametric devices that have one physical port for the processed signal and one resonance mode \cite{JPAimpedanceEng,StrongEnvCoupling,EngineeredFluxPumpedJPA,BroadbandFluxPumpedJPA,BroadbandCPWJPA,ImpedanceMatchingSNAILs,BroadbandSnakes}. 

In this work, we extend the impedance-matching design technique to nondegenerate Josephson mixers, that have two physical ports and two resonance modes and whose nonlinear medium, i.e., the JRM, has four main nodes, none of which can be shorted to ground. Furthermore, we incorporate into the new designs key theoretical and experimental lessons shown to enhance the saturation power of JMs \cite{JPCsathighorder,Multiparametric}, such as designing them to operate as close as possible to Kerr-nulling working points of the JRM, and realizing favorable inductance ratios between the inductances of the JRMs and the resonators they are embedded into \cite{Roch,OptJPC}. 

Furthermore, we implement the new JM designs using a trilayer Nb process typically used in the fabrication of TWPAs \cite{TWPAScience} and SQUIDs \cite{SQUIDworkIBM}, which enable us, among other things, to realize the required lumped-element capacitors and inductors. In particular, we realize and measure four JM devices of the new design and fabrication process. Two of them implement lumped-element resonator-based JMs, which we refer to as resonant-mode JMs, while the other two incorporate matching networks, which we refer to as coupled-mode JMs. We show that the redesigned lumped-element JM devices can operate near the quantum limit in amplification and frequency conversion. We demonstrate that a lumped-element, resonant-mode JM device with low external quality factor can achieve relatively large bandwidths and saturation powers, i.e., about $600$ MHz and above $-100$ dBm, when operated in conversion mode, which, unlike the amplification mode, does not follow the amplitude-gain bandwidth product, known to significantly limit the dynamical bandwidths of parametric amplifiers at high gains \cite{microstripJPC,JPCreview}. Also, importantly, we show that the coupled-mode JMs exhibit enhanced bandwidths and saturation powers in amplification (conversion) of about $400$ MHz ($700$ MHz) and $-110$ dBm ($-95$ dBm).
 
\section{Three-wave mixing with the JRM} 
 
An ideal nondegenerate three-wave mixing device absorbs three input signals, referred to as signal, idler, and pump with distinct angular frequencies $\omega_s$, $\omega_i$, and $\omega_p$, and complex amplitudes $A^{\rm{in}}_s$, $A^{\rm{in}}_i$, and $A^{\rm{in}}_p$, where $\omega_p$ corresponds to either the frequency sum $\omega_s+\omega_i$ or difference $|\omega_i-\omega_s|$ and emits output signals at the same frequencies with amplitudes $A^{\rm{out}}_s $, $A^{\rm{out}}_i$, and $A^{\rm{out}}_p$, which, without internal dissipation, satisfy the energy conservation condition $\left| A^{\rm{in}}_s\right|^2+\left| A^{\rm{in}}_i\right|^2+\left| A^{\rm{in}}_p\right|^2=\left| A^{\rm{out}}_s\right|^2+\left| A^{\rm{out}}_i\right|^2+\left| A^{\rm{out}}_p\right|^2$. 

When a strong coherent pump tone is applied, i.e., $\left| A^{\rm{in}}_p\right|^2\gg \left| A^{\rm{in}}_s\right|^2, \left| A^{\rm{in}}_i\right|^2$ such a device can operate in two modes depending on the pump frequency, (1) signal and idler amplification with photon gain when $\omega_p=\omega_s+\omega_i$, and (2) noiseless frequency conversion without photon gain between the signal and idler when $\omega_p=|\omega_i-\omega_i|$. While in the former case, the pump drive, which undergoes a downconversion process, supplies the extra photon numbers in the emitted signals, in the latter case, the pump drive provides the energy difference between the photons at $\omega_i$ and $\omega_s$.    

The JRM, which is the nonlinear element in nondegenerate JM devices, implements such dispersive three-wave mixing operation in the microwave domain. It consists of four Josephson junctions, each with Josephson inductance $L_{J0}=\varphi_0/I_0$, forming a small superconducting loop, i.e., $L_{J0}\gg L_s$, which is threaded by an external flux $\Phi_e$ and has the symmetry of a Wheatstone bridge as shown in Fig.\,\ref{Device}(a). Due to the JRM structure and symmetry, its four nodes couple to three orthogonal spatial modes, two that are differential $\varphi_a$ and $\varphi_b$ corresponding to the signal and idler, while the third $\varphi_c$ is a common excitation corresponding to the pump. Expressing the normal modes $\varphi_{a}$, $\varphi_{b}$, $\varphi_{c}$ in terms of the dimensionless node fluxes $\varphi_{1}$, $\varphi_{2}$, $\varphi_{3}$, $\varphi_{4}$ (shown in Fig.\,\ref{Device} (a)) gives $\varphi_{a}=\varphi_{1}-\varphi_{3}$, $\varphi_{b}=\varphi_{2}-\varphi_{4}$, and $\varphi_{c}=\varphi_{1}+\varphi_{3}-\varphi_{2}-\varphi_{4}$, where $\varphi_j(t)=\frac{1}{\varphi_0}\int^{t}_{-\infty}V(t')dt'$ for node $j$ and $\varphi_0=\hbar/2e$ is the reduced flux quantum.

Using the normal spatial modes, the energy of the JRM can be written as \cite{OptJPC}

\begin{align}
	E_{\rm{JRM}}=&-4E_J \cos\left( \dfrac{\varphi_a}{2}\right)\cos\left( \dfrac{\varphi_b}{2}\right)\cos\left( \dfrac{\varphi_c}{2}\right)\cos\left( \dfrac{\varphi_e}{4}\right)  \nonumber \\
	&-4E_J\sin\left( \dfrac{\varphi_a}{2}\right)\sin\left( \dfrac{\varphi_b}{2}\right)\sin\left( \dfrac{\varphi_c}{2}\right)\sin\left( \dfrac{\varphi_e}{4}\right) \nonumber \\
	&+\dfrac{E_L}{4}\left(\varphi^2_a+\varphi^2_b+\dfrac{\varphi^2_c}{2} \right), \label{E_JRM}
\end{align}  

\noindent where $E_J=\varphi_0I_0$ is the Josephson energy, $E_L=\varphi^2_0/L_{in}$ is the energy associated with each of the linear inductors $L_{in}$ inside the loop, and $\varphi_e=\Phi_e/\varphi_0$.

Expanding $E_{\rm{JRM}}$ around the ground state, i.e., $\varphi_a=\varphi_b=\varphi_c=0$, which we assume is stable as we vary the external flux bias, we get \cite{Roch}

\begin{align}
	E_{\rm{JRM}}=& -\frac{1}{2}E_J\sin\left( \dfrac{\varphi_e}{4}\right)\varphi_a\varphi_b\varphi_c \nonumber \\
	&+\left[\frac{E_L}{4}+\frac{E_J}{2}\cos\left( \frac{\varphi_e}{4}\right) \right] \left(\varphi^2_a+\varphi^2_b\right)  \nonumber \\
	&+\left[\frac{E_L}{8}+\frac{E_J}{2}\cos\left( \frac{\varphi_e}{4}\right) \right]\varphi^2_c+O\left(\varphi^4_{a,b,c} \right) . \label{E_JRM_approx}
\end{align}  

The first term in Eq.\,(\ref{E_JRM_approx}) is the trilinear coupling term that gives rise to the desired three-wave mixing operation, whereas all the other couplings can be detrimental to the ideal performance of the device. As seen in Eq.\,(\ref{E_JRM}), the nonlinear couplings are controlled by the external magnetic flux $\varphi_e$. The Kerr nulling point for which the even-order couplings set by the cosine terms in Eq.\,(\ref{E_JRM}) vanish, is attained at $\varphi_e=2\pi$ (corresponding to $\Phi_e=\Phi_0$, where $\Phi_0$ is the flux quantum). 

Furthermore, using the series expansion of the dimensionless potential energy 
$E^{\prime}_{\rm{JRM}}=E_{\rm{JRM}}/E_L$, we obtain the dimensionless unpumped three-mode coupling strength 

\begin{equation}
	g=\frac{1}{2\beta}\sin\left(\dfrac{\varphi_e}{4}\right), \label{Dimensionlessg}
\end{equation}

\noindent where the inductance ratio $\beta=L_{J0}/L_{in}$ sets the overall strength of the nonlinearity. 

Another design parameter that significantly influences the strength of the various coupling terms of the mixing process is the participation ratio $0<p\leq 1$, which represents the fraction of the mode power within the JM device that is carried by the JRM \cite{JPCreview,OptJPC,OptJPCQuant}. 

Generally, lowering $p$, e.g., by increasing the series inductance outside the ring $L_{out}$, decreases the JRM nonlinearity by suppressing all coupling terms, but the suppression is not uniform. The higher order coupling terms get reduced more than the lower order ones. Thus, in certain parameter space, primarily of $\beta$ and $p$ investigated in Ref. \cite{OptJPC}, decreasing $p$ can boost the JM saturation power.   

\section{The device circuit}

\begin{figure*}
	[tb]
	\begin{center}
		\includegraphics[
		width=2\columnwidth 
		]%
		{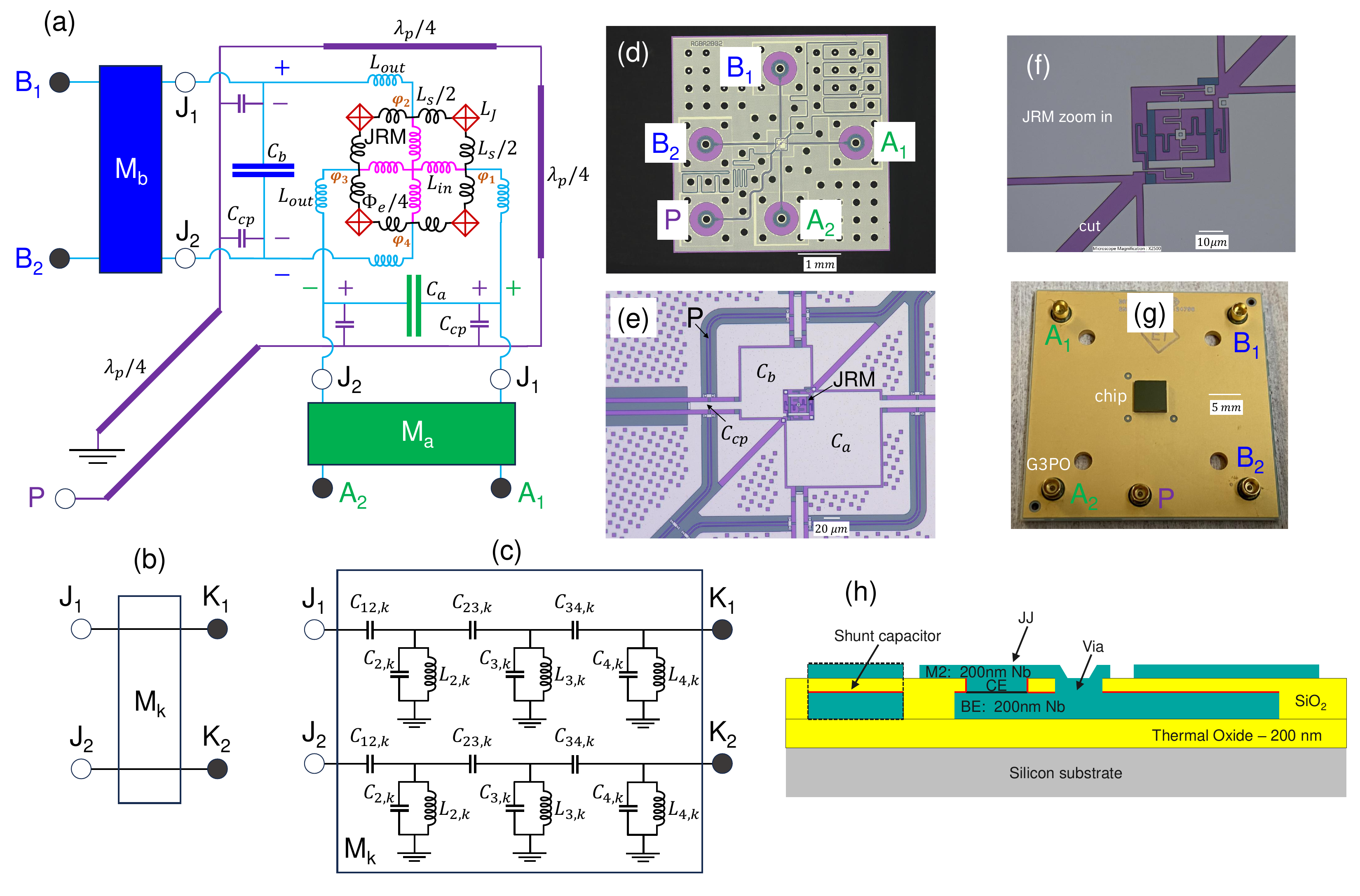}
		\caption{(a) Circuit diagram of the impedance-matched, lumped-element Josephson mixers. The capacitively-shunted JRM is connected to the device ports $\rm{A}_{1,2}$ and $\rm{B}_{1,2}$ through direct transmission lines for $\rm{JM_1}$ and $\rm{JM_2}$ as illustrated in (b), or through capacitively coupled three $LC$ resonators for $\rm{JM_3}$ and $\rm{JM_4}$ as shown in (c). (d) Optical image of a $\rm{JM_2}$ chip. The black disks on chip represent bump bonding sites. The launch pads for ports $\rm{A}$, $\rm{B}$, and $\rm{P}$ are located on the periphery. (e) Optical image of the capacitively shunted JRM. (f) A zoom-in on the JRM configuration incorporated in $\rm{JM_1}$ and $\rm{JM_2}$. (g) Image of a bump-bonded JM to a PCB and the G3PO connectors used for delivering the input and output signals. (h) Diagram of the multi-layer stack that implements three key elements of the JMs namely, plate capacitors, Nb Josephson junctions, and vias. In the diagram, M2 represents the second layer of Nb, BE and CE correspond to the bottom and counter electrodes of the Josephson junctions, which are separated by a thin layer of $\rm{Al/AlO_x}$ (represented by the black line), whereas the thin red line represents an anodized $\rm{NbO_x}$ layer.}
		\label{Device}
	\end{center}
\end{figure*}

The circuit elements of the realized JMs are shown in Fig.\,\ref{Device} (a). At the center, lies the JRM, which consists of four nominally identical Nb Josephson junctions connected in a superconducting loop. Each pair of opposite nodes is connected to linear outer inductors $L_{out}$ and shunted by plate capacitors $C_a$ and $C_b$, which together with the JRM inductance set the bare differential mode resonance frequencies for mode a and b. $L_s$ represents the geometric (stray) inductance associated with the superconducting leads connecting adjacent Josephson junctions. Inside the loop, the JRM junctions are shunted by linear inductors, i.e., $L_{in}$, which play two key roles, (1) they eliminate the hysteretic dependence of the JRM on the external flux threading the loop $\Phi_e$ \cite{Roch}, thus making the bare resonance frequencies of the device tunable with flux. Such flux-tunability is highly desirable and particularly crucial when frequency matching is required between various quantum devices \cite{QuantBAScience}, (2) they set the parameter $\beta$, which determines, as seen in Eq.\,(\ref{Dimensionlessg}), the  coupling parameter $g$ of the wave-mixing process that affects, among other figures, the saturation power of the JM and the maximum achievable gain in amplification or transmission in conversion.  

The blocks marked as $\rm{M_a}$ as $\rm{M_b}$, refer to impedance-matching networks, which connect the electrodes of the shunt capacitors $C_a$ and $C_b$ to the corresponding external ports of mode a and mode b, i.e., $\rm{A_{1,2}}$ and $\rm{B_{1,2}}$. In Fig.\,\ref{Device} (b) and (c) we depict the two circuit configurations for the $\rm{M_{k}}$ blocks employed in this work, where the index $\rm{k}$ refers to mode a or b. The configuration of panel (b), applicable to $\rm{JM_1}$ and $\rm{JM_2}$ devices (see Table \ref{JMroletype}), realizes direct leads (transmission lines) between node $\rm{J_i}$ and port $\rm{K_i}$. Whereas, the configuration of panel (c), applicable to $\rm{JM_3}$ and $\rm{JM_4}$ (see Table \ref{JMroletype}), realizes three alternating coupling capacitors and $LC$ resonators shorted to ground between ports $\rm{J_i}$ and $\rm{K_i}$.

\begin{table*}[tbh]
	\centering
	\begin{tabular}{c c c c c  c  c  c  c  c } 
		\hline
		Device & $\rm{JM_1}$ & $\rm{JM_2}$ & $\rm{JM_3}$ & $\rm{JM_4}$ \\ [0.5ex] 
		\hline
		Operation & Amplifier  & Transducer & Amplifier & Transducer   \\ 
		Mode design & Resonant & Resonant & Coupled & Coupled  \\
		Modes per port & 1 & 1 & 4 & 4  \\
			\hline 
	\end{tabular}
	\caption{The operation and configuration of the JM devices presented in this work.}
	\label{JMroletype}
\end{table*}

\begin{figure*}
	[tb]
	\begin{center}
		\includegraphics[
		width=2\columnwidth 
		]%
		{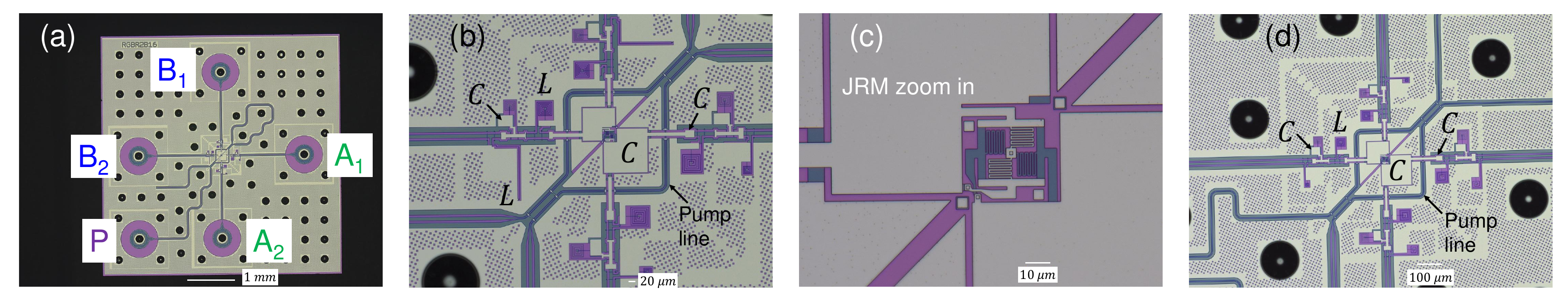}
		\caption{(a) Photo of a $\rm{JM_3}$ chip. (b) Zoom-in photo of a $\rm{JM_3}$ chip, showing the capacitively-shunted JRM at the center and the capacitively-coupled $LC$ resonators, which realize the lumped-element matching networks exhibited in Fig.\,\ref{Device} (c). (c) Optical image of the JRM incorporated in $\rm{JM_3}$ and $\rm{JM_4}$. (d) A similar image as (b) taken for a $\rm{JM_4}$ chip.          
		}
		\label{DeviceImage}
	\end{center}
\end{figure*}

The pump drive is fed to the JM through a separate port P connected to an on-chip transmission line as shown in panel (a). The pump transmission line is capacitively coupled via small plate capacitors $C_{cp}\simeq 60$ fF to the electrodes of the shunt capacitors $C_a$ and $C_b$, which are connected via $L_{out}$ to the four nodes of the JRM. To ensure that the pump drive excites the common mode of the JRM, i.e., opposite nodes are driven by the same polarity, we separate the coupling points to $C_a$ and $C_b$ by an on-chip transmission line of length $\lambda_p/2$, corresponding to a $\pi$ phase shift, where $\lambda_p$ refers to the mean pump wavelength. Furthermore, to ensure having rf-voltage anti-nodes near the locations of $C_{cp}$, we add a $\lambda_p/4$ section shorted to ground after the last $C_{cp}$.   
 
In Fig.\,\ref{Device} (d), we exhibit an optical image of a $\rm{JM_2}$ chip, which features the capacitively shunted JRM at the center, the bump-bonding sites (black circles), and the launching pads for ports $\rm{A_{1,2}}$, $\rm{B_{1,2}}$, and $\rm{P}$ on the periphery. A zoom-in image of the device center is shown in Fig.\,\ref{Device} (e). It displays the JRM, the shunt capacitors $C_{a,b}$, the pump transmission line surrounding $C_{a,b}$, and the pump coupling capacitors $C_{cp}$. A further zoom-in on the JRM is exhibited in Fig.\,\ref{Device} (f). The layout featured in the image is applicable to $\rm{JM_1}$ and $\rm{JM_2}$ devices. Superconducting vias are used to connect two of the outer inductors to the electrodes of $C_{a,b}$, and connect the four $L_{in}$ inside the JRM. A wire cross-over is formed between the bottom and left $L_{out}$, which are implemented using two different superconducting layers separated by an insulator. We estimate the parasitic capacitance introduced by the cross-over (which we neglect in our modeling) to be small of about $0.13$ fF ($0.5$ fF) in the case of $\rm{JM}_{1,2}$ ($\rm{JM}_{3,4}$). The design also includes a diagonal cut in the superconducting ground surrounding the JRM to suppress screening currents and thereby enable flux biasing of the JRM via a small superconducting coil mounted on top of the device cover. A photo of the JM chip bump bonded to a printed circuit board (PCB) with soldered G3PO connectors is shown in Fig.\,\ref{Device} (g). A diagram of the multi-layer stack used in the fabrication of key elements of the JMs, such as Nb Josephson junctions, plate capacitors, and vias, is shown in Fig.\,\ref{Device} (h).  

In Fig.\,\ref{DeviceImage}, we exhibit additional four optical images, showing $\rm{JM_3}$ and $\rm{JM_4}$ devices. Panel (a) exhibits a photo of a $\rm{JM_3}$ chip. Note that the on-chip pump line is much shorter in this instance than in Fig.\,\ref{Device} (d), since $\rm{JM_3}$ is designed and operated as an amplifier, which requires significantly shorter pump wavelengths corresponding to higher pump frequencies than $\rm{JM_2}$ operated as a frequency converter (e.g., $18$ GHz vs. $3$ GHz). Photos of $\rm{JM_3}$ and $\rm{JM_4}$ chips featuring their capacitively-shunted JRMs and capacitively-coupled $LC$ resonators are shown in (b) and (d), respectively. In these designs, plate capacitors are scaled based on their area, small inductors are realized as straight lines, while relatively large ones are realized as spirals. 

A photo of the JRM employed in the design of $\rm{JM_3}$ and $\rm{JM_4}$ devices is shown in Fig.\,\ref{DeviceImage} (c). The JRM configuration in this case differs from the one incorporated in $\rm{JM_1}$ and $\rm{JM_2}$ (shown in Fig.\,\ref{Device} (f)) in three main aspects: (1) the critical current of the Josephson junctions is smaller, i.e., $2.5$ $\mu$A vs. $8.1$ $\mu$A, (2) $L_{in}$ is larger, i.e., $30$ pH vs. $11.3$ pH, and (3) $L_{out}$ is smaller, i.e., $9$ pH vs. $27$ pH (see Tables \ref{JM12params}, \ref{JM3params}, \ref{JM4params}).

\begin{figure*}
	[tb]
	\begin{center}
		\includegraphics[
		width=2\columnwidth 
		]%
		{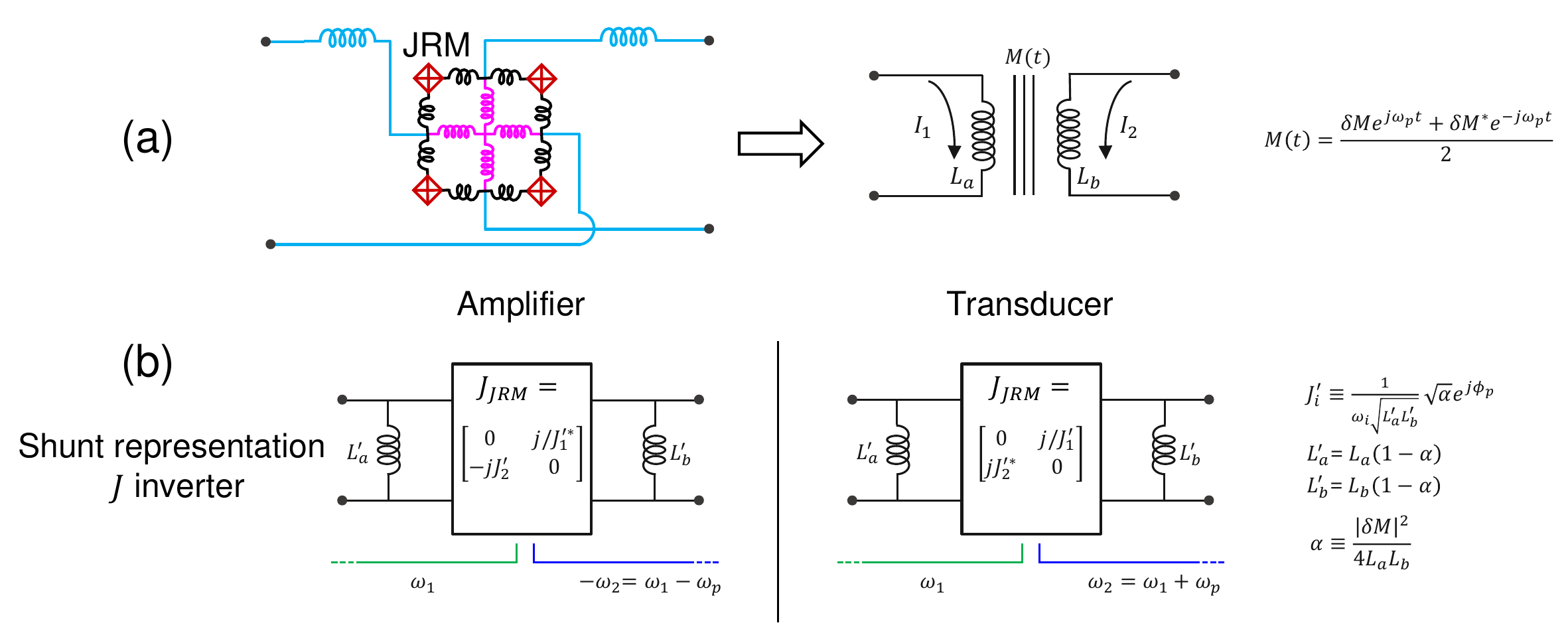}
		\caption{(a) When driven by a pump, the JRM operation can be modeled as parametric-modulation of a mutual inductance $M(t)$ between two linear inductors $L_a$ and $L_b$ with signal and idler currents $I_1$ and $I_2$. (b) Equivalent two-port circuit model  of the modulated mutual in the shunt representation corresponding to the  amplification (left) and transduction (right) modes of operation. The boxed elements represent admittance inverters with the corresponding $ABCD$ matrices $J_{\rm{JRM}}$.            
		}
		\label{JRMinverter}
	\end{center}
\end{figure*}

To model and design JMs, it is important to recognize the functional role played by the JRM. When driven by a pump tone at the frequency sum or difference between the nondegenerate idler (b) and signal (a) circuits connected to it, it acts as a parametric coupler mediating between them. At the circuit level, the nonlinear coupling introduced by the JRM can be modeled as a modulated mutual inductance as shown in Fig.\,\ref{JRMinverter} (a). In Fig.\,\ref{JRMinverter} (b), we present the equivalent two-port circuit model of the parametrically modulated mutual in the shunt representation corresponding to the amplification ($f_p=f_b+f_a$) and transduction ($f_p=|f_b-f_a|$) modes of operation, shown on the left- and right-hand side, respectively. One key advantage of employing JRMs to realize such parametric coupling over grounded SQUIDs for example is that with the former the coupling is purely parametric since it nulls the passive part of the coupling, i.e., $M_0=0$ (which is independent of the pump), thus providing same-frequency isolation between the signal and idler circuits.     
    
\section{resonant-mode Josephson mixers}    

In this section, we present measurements taken for the resonant-mode JM design, i.e., $\rm{JM_1}$ and $\rm{JM_2}$. Both $\rm{JM_1}$ and $\rm{JM_2}$ devices share the same design, layout, and fabrication process. The main difference between them is in their mode of operation as listed in Table \ref{JMroletype}, which leads to different on-chip pump lines, especially with respect to their total length, set by the pump frequency range applicable in each case. Images of the pump lines employed in $\rm{JM_1}$ and $\rm{JM_2}$ devices can be seen in Fig.\,\ref{DeviceImage} (a) (belonging to $\rm{JM_3}$) and Fig.\,\ref{Device} (d), respectively.
 
\begin{figure*}
	[tb]
	\begin{center}
		\includegraphics[
		width=2\columnwidth 
		]%
		{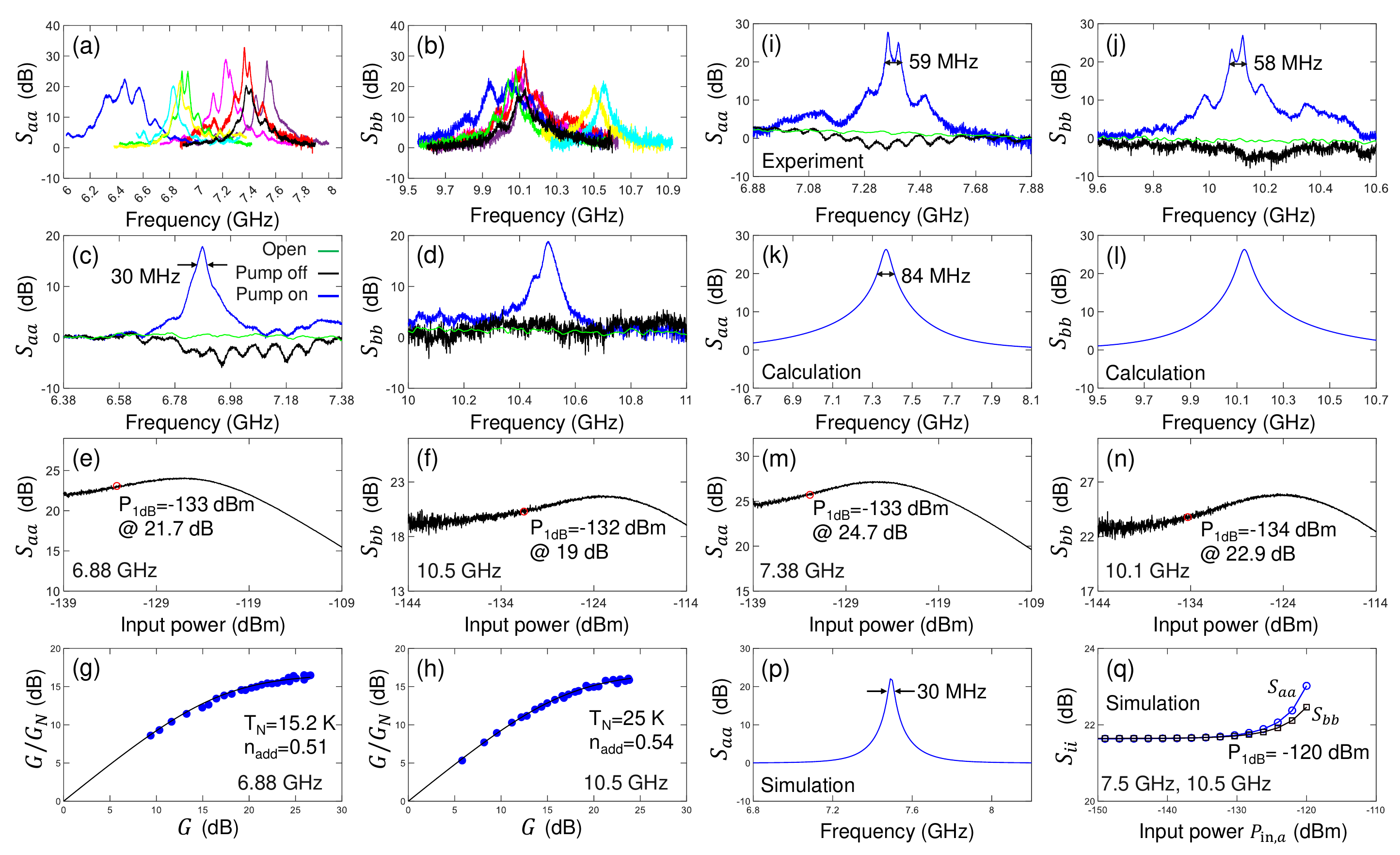}
		\caption{(a) and (b) reflection gain measurement results $S_{aa}$ and $S_{bb}$ of $\rm{JM_1}$ taken for different working points set by the applied flux, pump frequency, and power. Curves of the same color in (a) and (b) are taken at the same working point. (c) and (d) exhibit measured gain curves (blue) for one of the working points drawn in yellow in (a) and (b). The black curves correspond to the reflection parameters measured with pump off. The green curves, used as reference, are reflection parameters measured for open ends at nominally identical locations as the mounted devices. (e) and (f) saturation power measurements taken on resonance for the working point of (c) and (d). (g) and (h) SNR improvement for mode a and b measured versus the device gain (blue filled circles). The data is taken for the same flux and pump frequency as (c) and (d). The black curves represent theory fits. (i) and (j) same as (c) and (d) showing a different working point (red) exhibited in (a) and (b). (k) and (l) calculated response of the device that employs the extracted parameters listed in Table \ref{JM12params}. (m) and (n) saturation power measurements taken on resonance for the working point of (i) and (j). (p) Keysight ADS simulation of the reflection gain $S_{aa}$ versus frequency. (q) Keysight ADS simulation of $S_{aa}$ (blue circles) at $7.5$ GHz and $S_{bb}$ (black squares) at $10.5$ GHz versus the signal power applied to port a.        
		}
		\label{JM1wpt1}
	\end{center}
\end{figure*}

Based on the extracted design parameters of $\rm{JM_1}$ and $\rm{JM_2}$ (listed in Table \ref{JM12params}), we obtain $\beta\simeq3.6$ and $p\simeq 0.58$, where $p$ is evaluated using Eq.\,(\ref{p_ratio}). Moreover, since these JM devices consist of lumped-element parallel $LC$ resonators directly connected to the external feedlines of impedance $Z_0=50$ Ohm, their total quality factors are dominated by their external quality factors given by $Q_a=Z_0/Z_a\cong 14.8$ and $Q_b=Z_0/Z_b\cong 10.6$, where $Z_a=\sqrt{L_a/C_a}\cong 3.4$ Ohm, and  $Z_b=\sqrt{L_b/C_b}\cong 4.7$ Ohm (calculated for $\varphi_e=0$). Using these calculated values, we find that despite having low external quality factors both $\rm{JM_1}$ and $\rm{JM_2}$ devices satisfy the stability condition, known as the $pQ$ product, since $p^2Q_aQ_b\cong53\gg1$ \cite{microstripJPC,NatPhysJPC,JPCreview}.   

In Fig.\,\ref{JM1wpt1} we exhibit several key measurements of $\rm{JM_1}$. Plots (a) and (b), show reflection gain results $S_{aa}$ and $S_{bb}$ versus frequency measured for different working points, i.e., flux and pump parameters. The gain curves are normalized by the pump off response. Gain curves plotted with the same color in (a) and (b) are measured at the same working point. The measurement demonstrates that gains of $20$ dB or more can be obtained for modes a and b over a tunable bandwidth of about $1.1$ GHz and $600$ MHz, respectively. In (c) and (d) we plot in using blue color one of the pump-on working points shown in (a) and (b) in yellow, which exhibit about $20$ dB of gain and $30$ MHz of bandwidth. We also plot for reference the pump-off data (black) and reflection calibration data (green) taken for the same setup and conditions but with open-end terminations replacing the JM devices (see experimental setup shown in Fig.\,\ref{Setup}). By comparing the pump off data to the reference curves, we observe insertion loss of $0.5-3.5$ dB for mode a on resonance, and a negligible amount for mode b. Such a loss likely originates from dielectric loss in the plate capacitors. 

In plot (e) and (f), we exhibit saturation power measurements taken on resonance at the same working point as (c) and (d). In these measurements, we fix the pump parameters to yield gains of about $20$ dB (at sufficiently low signal power) and vary the input signal power applied to the JM at a certain frequency within its bandwidth. Saturation power indicated by red circles, i.e., $-133$ dBm in (e) and $-132$ dBm in (f), correspond to the minimum input power for which the gain changes by $\pm1$ dB relative to the low input power value. 

To show that $\rm{JM_1}$ amplifies near the quantum limit, we measure the SNR improvement of the output chain $G/G_N$ versus the JM gain on resonance for modes a and b shown in subplots (g) and (h), respectively. By using Eq.\,(\ref{SNR_improv}) to fit the data (black curve) we obtain estimates for $T_{\rm{N}}$ and $n_{\rm{add}}$, listed in the plot, which correspond to the effective noise temperature of the output chain and the noise-equivalent input photon number added by the JM. 

Similar to subplots (c) and (d), panels (i) and (j), exhibit data taken for another working point originally shown in (a) and (b) in red, which exhibit maximum gains of about $26$ dB and bandwidths of about $58$ MHz at $20$ dB. The pump-on gain curves (blue) are shown alongside the pump-off response (black) and the calibration reference (green). Like (c), the pump off response of the JM exhibits $1-3.5$ dB of loss on resonance relative to the reference level. 

As seen in Fig.\,\ref{JM1wpt1}, the measured gain curves generally have Lorentzian-like shapes and follow the amplitude-gain bandwidth product, which can be expressed as $\sqrt{G}B=\gamma$, where $B$ is the dynamical bandwidth of the JM, $\gamma=2\gamma_a\gamma_b/(\gamma_a\gamma_b)$ is the effective linear bandwidth of the JM, while $\gamma_{a}$ and $\gamma_{b}$ are the linear bandwidths of mode a and b, respectively. Based on the measured bandwidths $\gamma_{a}/2\pi\cong 400$ MHz and $\gamma_{b}\cong 1$ GHz for the device, we get $B/2\pi\simeq 57$ MHz at power gain of $20$ dB (i.e., $G=100$), which is close to the measured values in  Figs.\,\ref{JM1wpt1}(c),(d),(i),(j).   

Using the transfer matrix model of the JM described in Appendix A and B and the device parameters listed in Table \ref{JM12params} (extracted from the fits to the resonance frequency data versus flux, depicted in Fig.\,\ref{JM12freqvsflux} (a),(b)), we plot in panels (k) and (l) the calculated gain curves versus frequency of modes a and b which use the ideal lumped-element circuit of Fig.\,\ref{Device} (a), (b). 

Similar to the  measurements of (e) and (f), (m) and (n) exhibit saturation power measurement results for mode a and b taken on resonance at the working point of (i) and (j), which yield comparable saturation powers to (e) and (f), though obtained for about $3$ dB higher power gains than (e) and (f). It is worth noting that in both cases (e),(f) and (m),(n), the saturation power measurement results exhibit a ``shark-fin" feature, characterized by an initial rise in the gain with the input power followed by a monotonic decline. As shown in Refs. \cite{JPCsathighorder,OptJPC}, such a feature can be caused due to intrinsic self-Kerr or cross-Kerr effects (fourth-order coupling terms) that are dominant at the applied working point, or due to effective Kerr nonlinearity generated by a soft pump with only third-order couplings \cite{OptJPC}. While the latter can be caused when the pump drive is partially in resonance with the device, the former can arise for example when operating away from the Kerr-nulling point, which, importantly, does not null both self-Kerr and cross-Kerr terms simultaneously when $p<1$ \cite{OptJPC}. 

Furthermore, similar to the calculated gain response of $\rm{JM_1}$ presented in (k) and (l), we plot a microwave simulation result of $S_{aa}$ versus frequency in panel (p). The simulation is conducted using harmonic-balance in Keysight ADS ($2023$ version) based on the ideal lumped-element model of the JM depicted in Fig.\,\ref{Device} (a),(b). In Fig.\,\ref{JM1wpt1} (q), we plot the simulated results for $S_{aa}$ and $S_{bb}$ versus the input power applied to port a of the JM, i.e., $P_{\rm{in},a}$, obtained on resonance at the same working point of panel (p). Further details regarding the simulation of JM devices are discussed in Appendix D.

While the simulation result yields a saturation power that is about $13$ dB higher than in the experiment ($-120$ dBm vs. $-133$ dBm), the simulation result seems to capture the ``shark-fin" feature exhibited by the data. In this context, it is important emphasize that the saturation power values strongly depend on the exact working point parameters, such as the flux bias, the pump frequency detuning relative to the sum of the resonance frequencies of mode a and b (which is more challenging to determine in wideband devices), and the pump power. For instance, it has been shown in Ref. \cite{JPCsathighorder} that pump frequency detuning can influence the saturation power of a JM device by as much as $10$ dB.  

\begin{figure*}
	[tb]
	\begin{center}
		\includegraphics[
		width=2\columnwidth 
		]%
		{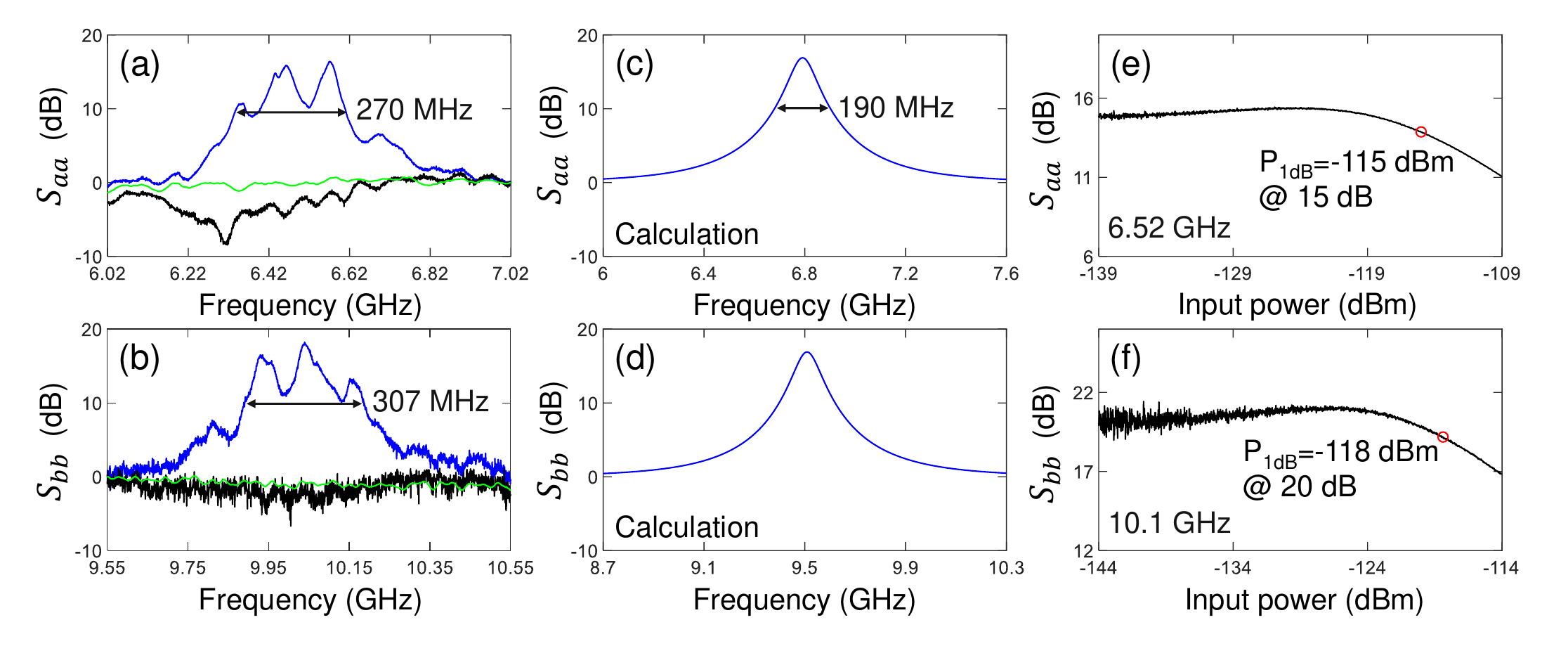}
		\caption{(a) and (b) exhibit measured gain curves (blue) of another working point of $\rm{JM_1}$ (shown in Fig.\,\ref{JM1wpt1} (a) and (b) in blue). Similar to Fig.\,\ref{JM1wpt1} (c) and (d) the black curves correspond to the reflection parameters measured with pump off and the green reference curves are reflection parameters measured for open ends at nominally identical locations as the mounted devices. The frequency range listed on the plots represent a bandwidth proxy taken at $10$ dB. (c) and (d) Calculated response of $\rm{JM_1}$ that employs the extracted parameters listed in Table \ref{JM12params}. (e) and (f) saturation power measurements taken at $6.52$ GHz ($S_{aa}$) and $10.1$ GHz ($S_{bb}$) for the working point of (a) and (b). Notably, the saturation powers obtained at this working point is about $14-19$ dB higher than those measured in Fig.\,\ref{JM1wpt1} (e), (f), (m), (n).           
		}
		\label{JM1wpt2}
	\end{center}
\end{figure*}

Another noteworthy working point of $\rm{JM_1}$, whose gain curves show pronounced deviation from the Lorentzain-like shape, is displayed in Fig.\,\ref{JM1wpt2} (originally drawn in blue in Fig.\,\ref{JM1wpt1} (a) and (b)). As seen in Fig.\,\ref{JM1wpt2} (a) and (b), the measured gain curves $S_{aa}$ and $S_{bb}$ exhibit bandwidths of about $270$ MHz and $307$ MHz at $10$ dB of gain, which are significantly larger than those obtained for the other working points presented in Fig.\,\ref{JM1wpt1}. For comparison, we plot in Fig.\,\ref{JM1wpt2} (c) and (d), the calculated gain curves $S_{aa}$ and $S_{bb}$ versus frequency near the flux bias point applied in the experiment. Furthermore, as shown in Fig.\,\ref{JM1wpt2} (e) and (f), the saturation powers achieved in this case, i.e., $-115$ dBm and $-118$ dBM, are at least $14$ dB higher than what is measured for the other two working points in Fig.\,\ref{JM1wpt1} of about $-132$ dBm.  

One possible explanation for the significantly enhanced saturation power achieved in Fig.\,\ref{JM1wpt2} compared to Fig.\,\ref{JM1wpt1}, is that the flux bias $\Phi_e=1.6\Phi_0$ in the measurement of Fig.\,\ref{JM1wpt2} is closer to a Kerr-nulling point for the device compared to $\Phi_e=1.24\Phi_0$ in Fig.\,\ref{JM1wpt1} (m),(n) (see Table VI). Such a deviation of the Kerr-nulling point from the $\Phi_e=\Phi_0$ value, can be explained by the presence of $L_s\neq 0$ in our device, which is known to shift the Kerr-nulling point upwards as $L_s$ is increased. Moreover, based on a calculation in Ref. \cite{Multiparametric} (done for $\beta=4.5$) the nulling point shifts to about $1.55\Phi_0$ for an inductance ratio $\alpha_J=L_s/L_{J0}\simeq 0.25$ applicable to $\rm{JM}_1$.

\begin{figure*}
	[tb]
	\begin{center}
		\includegraphics[
		width=2\columnwidth 
		]%
		{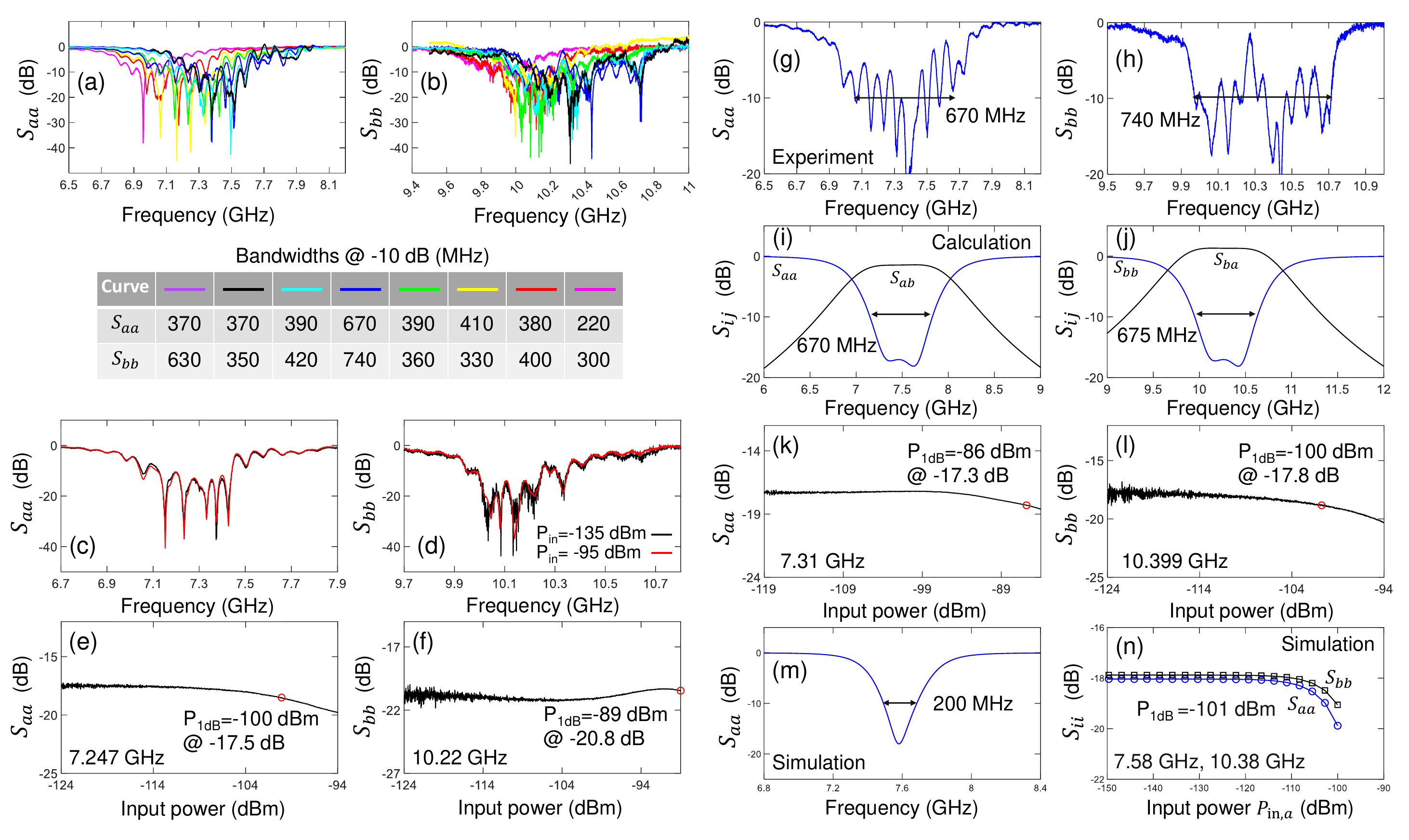}
		\caption{(a) and (b) reflection magnitude measurement results $S_{aa}$ and $S_{bb}$ of $\rm{JM_2}$ taken for different working points set by the applied flux, pump frequency, and power. Curves of the same color in (a) and (b) are taken at the same working point. The table beneath (a) and (b) lists a bandwidth proxy of the various working points for mode a and b taken at $-10$ dB. The values in the table are rounded down to the nearest tens of Megahertz. (c) and (d) exhibit measured reflection magnitude curves (black) for one of the working points drawn in green in (a) and (b). The red curves which mostly overlap with the black curves correspond to the same working point measured at higher input power, i.e., $-95$ dBm versus $-135$ dBm. (e) and (f) saturation power measurements taken at $7.247$ GHz ($S_{aa}$) and $10.22$ GHz ($S_{bb}$) for the working point of (c) and (d). (g) and (h) exhibit measured reflection magnitude curves for one of the working points drawn in blue in (a) and (b) (with the largest bandwidth). The frequency range listed on the plots represent a bandwidth proxy taken at $-10$ dB. (i) and (j) calculated scattering parameters of the device, $S_{aa}$ and $S_{bb}$ in blue and $S_{ab}$ and $S_{ba}$ in black. The theoretical model used in the calculation employs the extracted parameters listed in Table \ref{JM12params}. (k) and (l) saturation power measurements taken at $7.31$ GHz ($S_{aa}$) and $10.399$ GHz ($S_{bb}$) for the working point of (g) and (h). (m) Keysight ADS simulation of the reflection magnitude $S_{aa}$ versus frequency. (n) Keysight ADS simulation of $S_{aa}$ (blue circles) at $7.58$ GHz and $S_{bb}$ (black squares) at $10.38$ GHz versus the signal power applied to port a.     
		}
		\label{JM2wpt1}
	\end{center}
\end{figure*}

Similar to the tunable bandwidth measurement of $\rm{JM_1}$ shown in Fig.\,\ref{JM1wpt1} (a) and (b), we plot in Fig.\,\ref{JM2wpt1} (a) and (b), the reflection amplitudes $S_{aa}$ and $S_{bb}$ versus frequency measured for multiple working points of $\rm{JM_2}$. The measured curves are normalized by the pump off data. Same-color curves in (a) and (b) are taken at the same working point. As seen in the plots both modes exhibit a tunable range of about $600$ MHz. In the table beneath panels (a) and (b), we list a bandwidth proxy for each of the working points, which we adopt due to the relatively large ripples in the measured reflection response, which are caused by standing waves forming in the coaxial lines between the circulators and the device (see setup diagram in Fig.\,\ref{Setup} and Appendix G for more details). The proxy bandwidth values listed in the table are rounded down to the nearest tens of megahertz and correspond to the frequency span for which the reflection amplitude curves lie primarily below $-10$ dB and roughly cover the range in which appreciable frequency conversion is taking place in the JM. In Fig.\,\ref{JM2wpt1} (c) and (d), we contrast 
the frequency response of $S_{aa}$ and $S_{bb}$ measured for the same working point (green in (a) and (b)) using two different input powers, i.e., $-135$ dBm (black) versus $-95$ dBm (red). As seen in the figure, the two measured frequency responses of the JM exhibit near perfect overlap despite applying a $40$ dB higher input power than in the low power case. To further quantify the saturation power of the device at this working point, we exhibit in Fig.\,\ref{JM2wpt1} (e) and (f) saturation power measurements taken at $7.247$ GHz and $10.22$ GHz, respectively, which yield maximum input powers of $-100$ dBm for port a and at least $-89$ dBm for port b. 

The considerably higher saturation powers (by $20-40$ dB) measured in $\rm{JM}_2$ compared to $\rm{JM}_1$ can be attributed, to a large extent, to their different mode of operation, frequency conversion without photon gain versus amplification and also to possibly having a much stiffer pump in the conversion case, as their applied pump frequencies are vastly different, i.e., around $3$ GHz in conversion versus $17$ GHz in amplification.

In Fig.\,\ref{JM2wpt1} (g) and (h), we plot another working point of $\rm{JM_2}$ (drawn in blue in panels (a) and (b)), which yields the largest effective bandwidth of about $670$ MHz and $740$ MHz, respectively. Using the transfer matrix model of the JM and the device parameters listed in Table \ref{JM12params}, we calculate the reflection ($S_{aa}$, $S_{bb}$) and conversion parameters versus frequency ($S_{ab}$, $S_{ba}$) plotted in Fig.\,\ref{JM2wpt1} (i) and (j) using blue and black curves, respectively. As seen in the figure, we obtain an effective  bandwidth of about $670$ MHz, which is similar to the experimental value. Note that the different maximum levels of the calculated transmission parameters between $S_{ba}$ ($1.29$ dB) and $S_{ab}$ ($-1.45$ dB) originate from the Manely-Rowe relations \cite{ManelyRowe,Pozar}, requiring power gain (reduction) in upconversion (downconversion) processes that have unity photon gain, which correspond to the ratio of the output to the input frequency $10\log_{10}\left( \omega_b/\omega_a\right)=1.38$ dB ($-1.38$ dB). 

In Figs.\,\ref{JM2wpt1} (k) and (l) we exhibit the corresponding saturation power measurements for the working point of (g) and (h), which yield at about $-17$ dB for mode a and b, $-86$ dBm and $-100$ dBm, respectively. The simulated response $S_{aa}$ versus frequency near this working point is shown in Fig.\,\ref{JM2wpt1} (m). While the simulation result exhibits a considerably narrower bandwidth than in the experiment, i.e., $200$ MHz versus $670$ MHz, the simulated saturation power shown in Fig.\,\ref{JM2wpt1} (n) yields a fair agreement with the experimental value for mode b.  

One interesting aspect of the measurement results shown in Figs.\,\ref{JM1wpt1} and \ref{JM2wpt1}, is that despite having the same design, $\rm{JM_1}$ and $\rm{JM_2}$ exhibit large disparity in their typical measured bandwidths, i.e., tens versus hundreds of megahertz, which stem from their different mode of operation, i.e., amplification versus conversion. While in amplification, the JM follows the amplitude-gain bandwidth product limit, which results in decreased bandwidth with gain, in conversion it follows a similar limit for only one possible bandwidth metric. To elucidate this key distinction, we plot in Fig.
\,\ref{ConvBW}(a) a calculated response of an ideal resonant-mode JM operated in conversion (see Appendix F). The plot exhibits the reflection (red) and frequency-converted transmission (blue) amplitudes versus normalized frequency deviation, where $f_0$ is the resonance frequency of mode a or b of the JM. As illustrated in the plot, it is possible to define three bandwidth metrics: (1) frequency spacing of the reflection parameter at $+3$ dB (or any desired +X dB) above the on-resonance minimum, (2) frequency spacing of the reflection parameter at $-3$ dB (or any desired -X dB) below the off-resonance $0$ dB level, and (3) frequency spacing of the transmission parameter at $-3$ dB (or any desired -X dB) below the on-resonance maximum. 

To demonstrate how the shapes of the reflection and transmission curves change with pump power, we plot in Fig.\,\ref{ConvBW}(b) the calculated JM response for varying pump amplitudes. As seen in the plot, the transmission and reflection curves calculated using Eq.\,(\ref{Sba_param_vs_freq}) and Eq.\,(\ref{r_a_param_vs_freq}), become wider with the increased pump, while the reflection minimum becomes smaller and narrower. To quantify this effect, we plot in Fig.\,\ref{ConvBW}(c), the three extracted bandwidths based on the three metrics outlined above as a function of the normalized pump power $\rho^2$. As expected, the red curve associated with the reflection minimum decreases with the pump, while the black and blue curves associated with reflection and transmission maxima increase with pump and converge at a value that is close to the larger linear bandwidth of modes a and b. We also plot for reference in Fig.\,\ref{ConvBW}(d), the corresponding maximum transmission (blue) and minimum reflection (red) parameters.

\begin{figure}
	[tb]
	\begin{center}
		\includegraphics[
		width=\columnwidth 
		]%
		{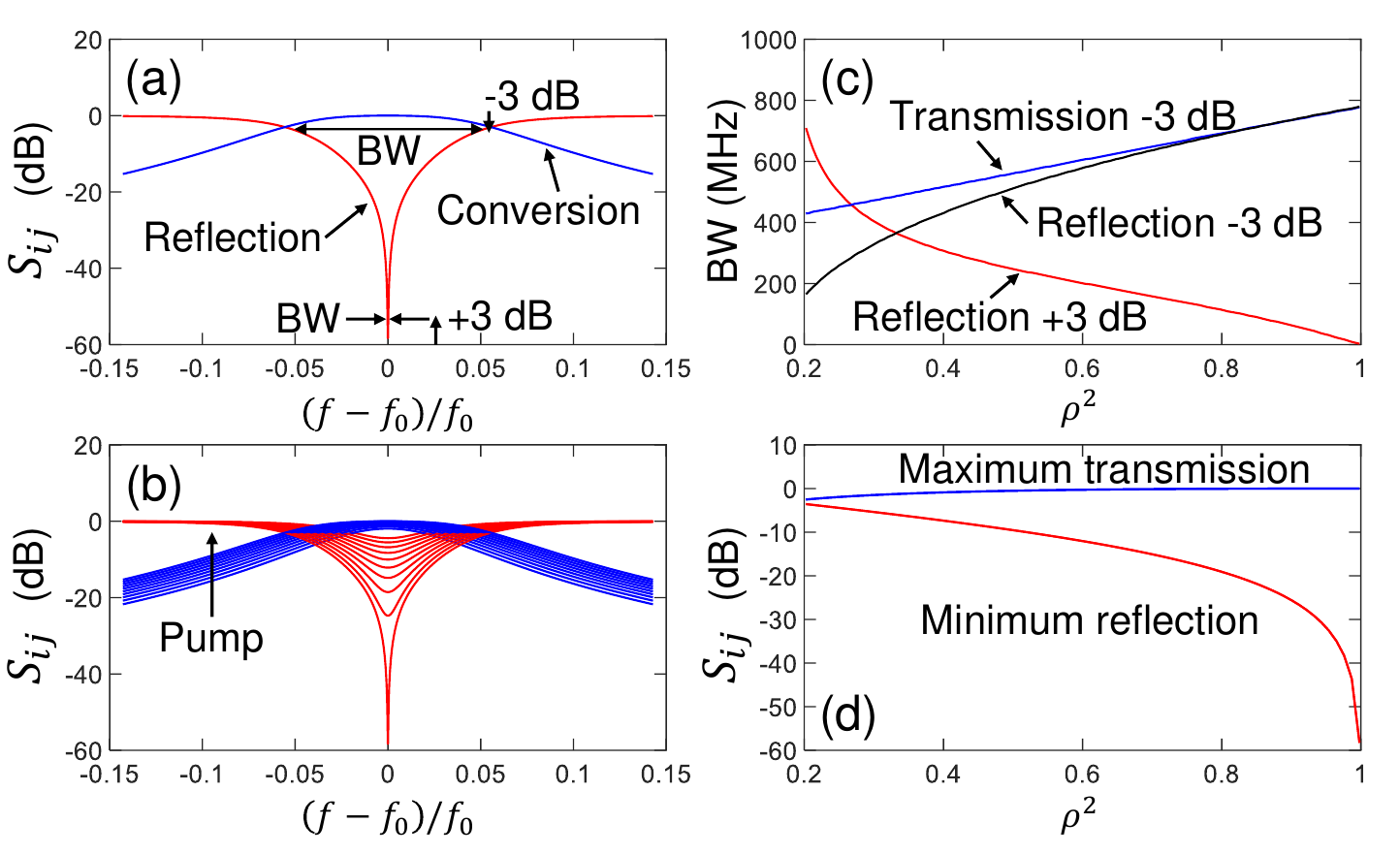}
		\caption{(a) Amplitudes of reflection (red) and transmission with frequency conversion (blue) of an ideal resonant-mode JM plotted versus normalized frequency. The two curves are calculated using Eq.\,(\ref{r_a_param_vs_freq}) and Eq.\,(\ref{Sab_param_vs_freq}). The plot serves to illustrate three possible bandwidth figures: (1) frequency difference between the $-3$ dB points of the reflection curve relative to the $0$ dB off-resonance level, (2) frequency difference between the $-3$ dB points of the conversion curve relative to the maximum transmission on resonance, and (3) frequency difference between the $+3$ dB points relative to the minimum reflection on resonance. (b) same as (a) calculated for an ascending normalized pump power. (c) displays the dependence of the three possible bandwidth figures illustrated in (a) as a function of normalized pump power $\rho^2$. (d) plots the corresponding maximum transmission with frequency conversion (blue) and minimum reflection (red) as a function of normalized pump power $\rho^2$. In the calculation, we employ $f_a=7$ GHz, $f_b=10$ GHz, $\gamma_a/2\pi=400$ MHz, $\gamma_b/2\pi=900$ MHz.          
		}
		\label{ConvBW}
	\end{center}
\end{figure}

\section{Coupled-mode Josephson mixers} 

In this section, we present measurement results taken for the coupled-mode JM design, i.e., $\rm{JM_3}$ and $\rm{JM_4}$, which are operated in amplification and frequency-conversion, respectively. 

To enhance the bandwidths of JM devices, especially of parametric amplifiers which are inherently limited by the amplitude-gain bandwidth product, we apply the versatile and quite effective technique for shaping the frequency response of microwave devices known as impedance matching, which can be realized by incorporating coupled-mode networks between the parametric nonlinear element and the external feedlines.

\begin{figure*}
	[tb]
	\begin{center}
		\includegraphics[
		width=2\columnwidth 
		]%
		{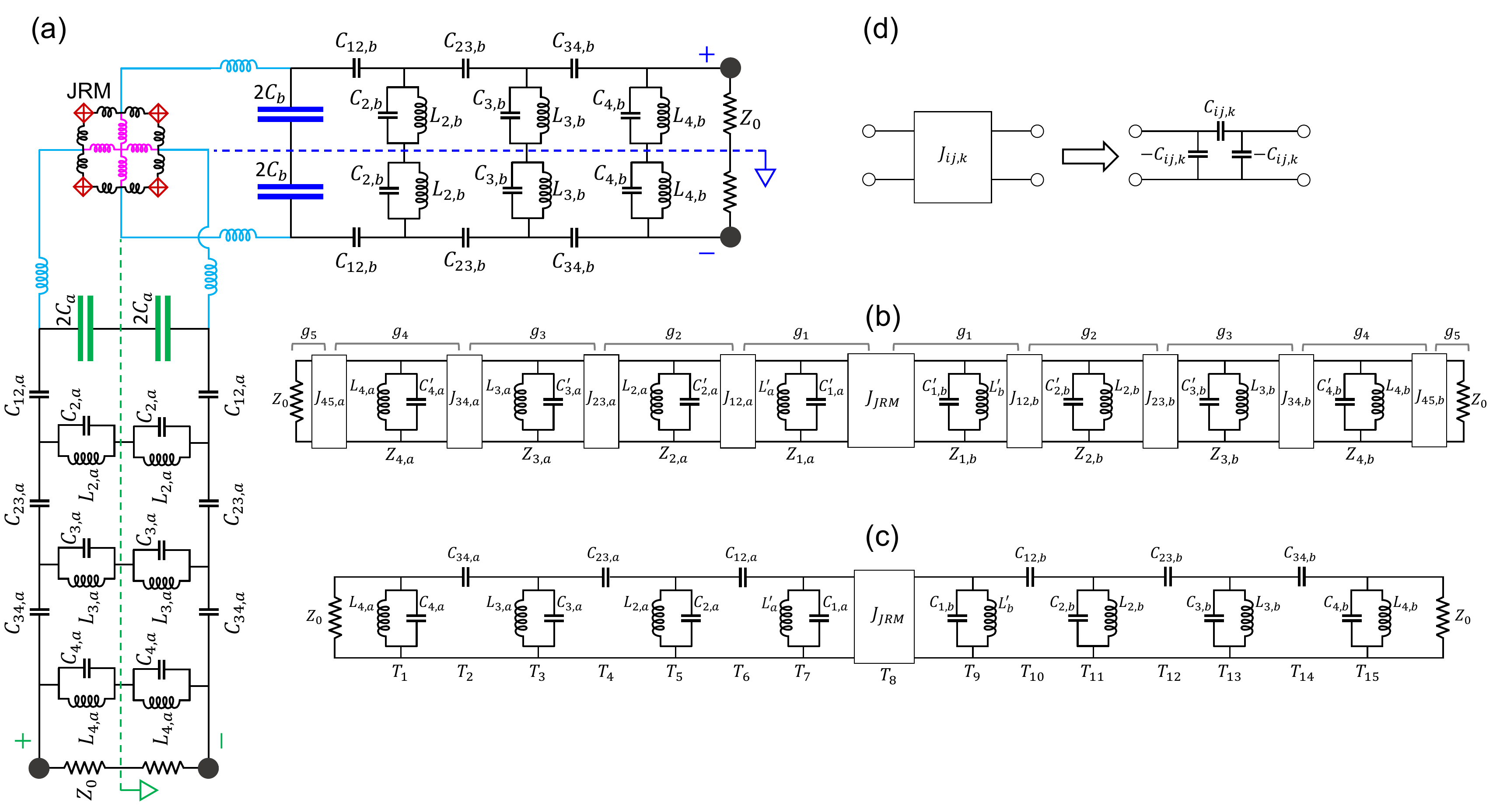}
		\caption{(a) Circuit model of the Josephson mixers with lumped-element, impedance matching networks, i.e., $\rm{JM_3}$ and $\rm{JM_4}$. In this model, we  replaced the single capacitor shunting opposite nodes of the JRM (shown in Fig.\,\ref{DeviceImage} (b) and (d)) with an equivalent fictitious two capacitors in series with double the capacitance, and joined the respective $LC$ resonators shorted to ground between ports $\rm{A_1}$ and $\rm{A_2}$ and $\rm{B_1}$ and $\rm{B_2}$ via a common virtual ground that passes through the symmetry axes of the differential modes of the JRM (indicated by dashed green and blue lines). (b) exhibits a circuit model in which half of the JRM represented as an admittiance inverter $J_{\rm{JRM}}$ is matched to the external feedlines via a series of parallel $LC$ resonators and admittance inverters $J_{ij,k}$. (c) forms the basis for (a) and it is derived from (b) after substituting the generic admittance inverters $J_{ij,k}$ with their corresponding lumped-element capacitive $\pi$ sections illustrated in (d) and absorbing any negative capacitance associated with these sections into the shunt capacitors of the $LC$ resonators.  
		}
		\label{CircuitModel}
	\end{center}
\end{figure*}

While there is a large number of useful references that apply impedance matching methods in a wide variety of microwave applications \cite{Pozar,StudyofbroadbandPA,PrototypeBroadPA,MethodWidebandPA,SynInterstageNetworks}, here we apply the design technique of coupled-mode networks outlined and distilled in Ref.\,\cite{SynParamCouplNet} for Josephson junction-based parametric devices. However, since the emphasis in that reference is on grounded SQUID-based parametric devices, we extend the prescribed engineering procedure for multi-node Josephson element such as the JRM, which cannot be grounded without sacrificing its critical ability to support nondegenerate modes.        
 
In Fig.\,\ref{CircuitModel} we outline the design scheme that we apply to the JRM case. To better explain the design process, we start from the final coupled-mode circuit design depicted in Fig.\,\ref{CircuitModel} (a). In addition to the fundamental modes of the device formed by shunting the JRM and the series outer inductors by lumped-element capacitors, each pair of differential ports a and b incorporates three capacitively-coupled parallel lumped-element $LC$ resonators. This circuit diagram represents an idealized version of $\rm{JM_3}$ and $\rm{JM_4}$ layout shown in Fig.\,\ref{DeviceImage} (b) and (d) with two main differences that are introduced for pedagogical purposes. First, we substitute the single shunt capacitor on each side of the JRM with two capacitors in series with double the capacitance. Second, we join the respective grounded $LC$ resonators on each arm of port a and b via their common ground connection. Such  changes highlight the structural symmetry that exists in the device including the JRM with respect to the virtual ground that passes through the middle of each pair of differential ports as indicated by the dashed blue and green lines. By taking half of the circuit extending along the virtual ground of both modes shown in Fig.\,\ref{CircuitModel} (b), we can cast the coupled-mode network of the JM in the standard grounded form used in Ref.\,\cite{SynParamCouplNet}, which incorporates admittance inverters $J_{ij,k}$ between resonators $i$ and $j$ belonging to mode $k$. Each resonator $i$ of mode $k$ is characterized by an impedance $Z_{i,k}$. The values of $J_{ij,k}$ are determined by the prototype coefficients of the filter design $g_i$ and $g_j$ and the impedances of the adjacent resonators (see Appendix C.1). After replacing the generic admittance inverters in Fig.\,\ref{CircuitModel} (b) with a circuit realization in the form of lumped-element capacitive $\pi$ sections illustrated in Fig.\,\ref{CircuitModel} (d), we obtain the desired circuit design parameters outlined in (c) (which form the basis of the final design presented in (a)). 

\begin{figure}
	[tb]
	\begin{center}
		\includegraphics[
		width=\columnwidth 
		]%
		{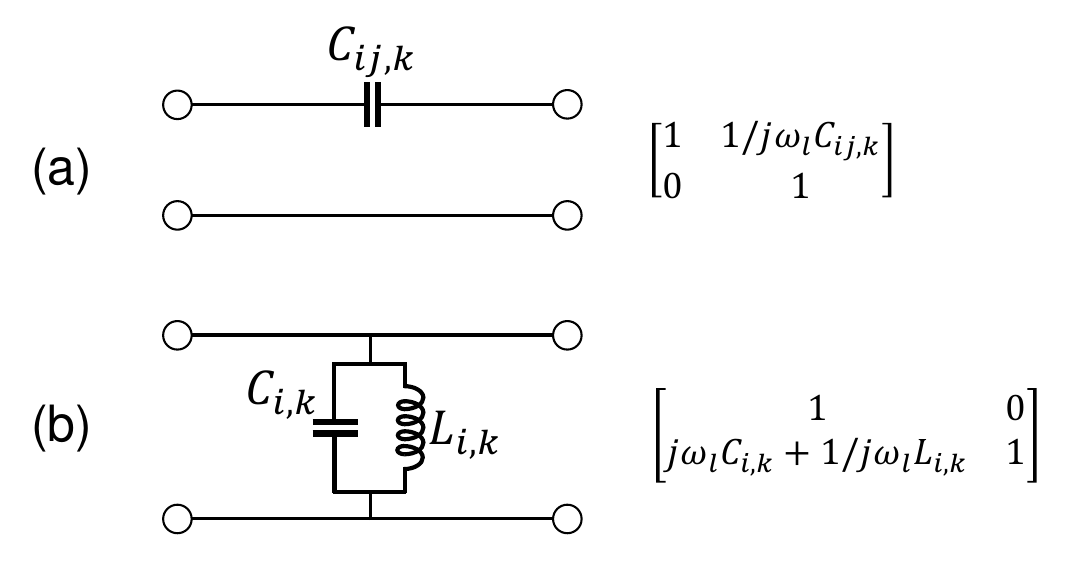}
		\caption{The corresponding $ABCD$ ($T$) matrices for a series capacitor $C_{ij,k}$ coupling stages $i$ and $j$ of mode $k$ (a) and $LC$ resonator $i$ of mode $k$ (b).  
		}
		\label{TransferMatrix}
	\end{center}
\end{figure}

To calculate the JM response, we multiply the $ABCD$ matrices of the various elements of the circuit represented by $T_i$. The $ABCD$ matrices of the coupling capacitor and parallel $LC$ resonators are shown in Fig.\,\ref{TransferMatrix}, whereas that of the JRM, which depends on the JM mode of operation, is shown in Fig.\,\ref{JRMinverter}.    
 
\begin{figure*}
	[tb]
	\begin{center}
		\includegraphics[
		width=2\columnwidth 
		]%
		{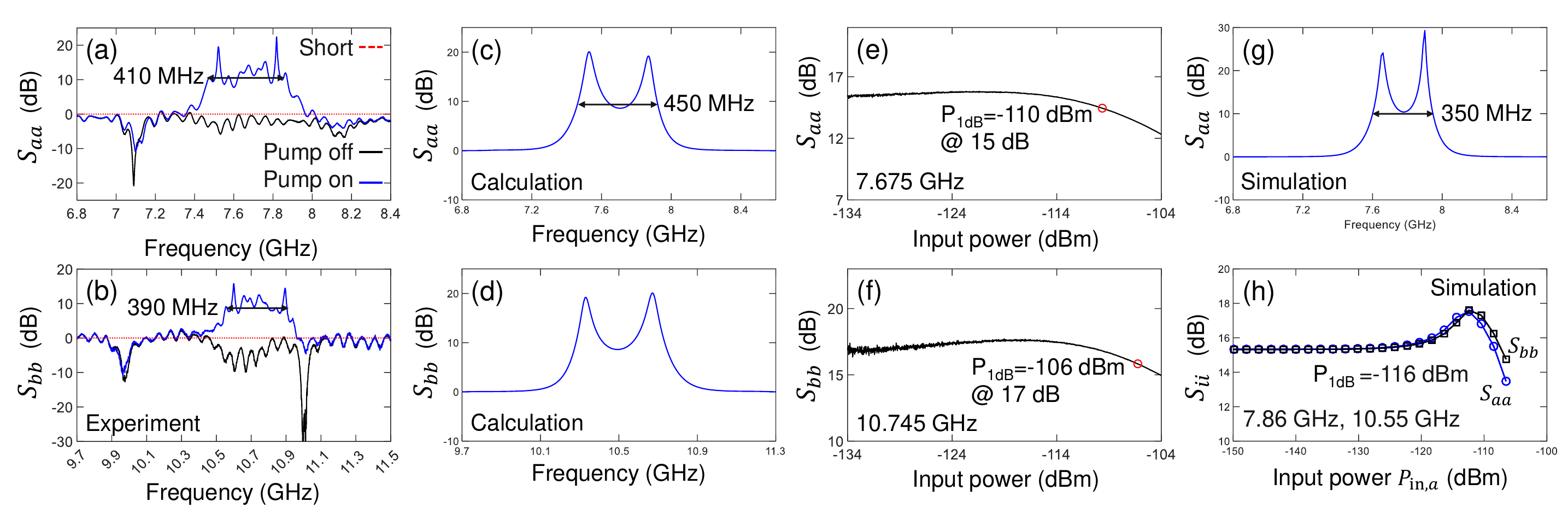}
		\caption{(a) and (b) reflection gain curves $S_{aa}$ and $S_{bb}$ (blue) of $\rm{JM}_3$ measured versus frequency (pump on). The black curves are reflection magnitude for mode a and b measured versus frequency with pump off. Both results are calibrated relative to the reflection magnitude measured for a short ended setup plotted as a $0$ dB guide-to-the-eye dashed red line. The frequency range listed on the plots represent a bandwidth proxy taken at $10$ dB. (c) and (d) calculated gain versus frequency for mode a and b. The theoretical model employs the extracted parameters of the device listed in Table \ref{JM3params}. (e) and (f) saturation power measurements taken at $7.675$ GHz and $10.745$ GHz for the working point of (a) and (b). (g) Keysight ADS simulation of the reflection gain $S_{aa}$ versus frequency. (h) Keysight ADS simulation of $S_{aa}$ (blue circles) at $7.86$ GHz and $S_{bb}$ (black squares) at $10.55$ GHz versus the signal power applied to port a.         
		}
		\label{JM3wpt1}
	\end{center}
\end{figure*}

In Fig.\,\ref{JM3wpt1} (a) and (b), we exhibit $S_{aa}$ and $S_{bb}$ of $\rm{JM_3}$ measured versus frequency for pump on (blue) and off (black). Both measurement results are normalized relative to the calibrated reflection data measured in-situ for a short termination (red). As seen in the figure, the amplification gain measured on ports a and b exceed $10$ dB within a bandwidth of $410$ MHz and $390$ MHz, respectively. Amplitude dips observed in the device response outside the device bandwidth could originate from unmatched fabricated values of one or two $LC$ resonators along the symmetry lines of the device (shown in Fig.\,\ref{CircuitModel} (a)), causing some of the energy of the excited differential mode at the dip frequencies to leak into a common mode, which gets absorbed by the cold $50$ Ohm termination attached to the $\Sigma$ port of the hybrid employed in the experimental setup (Fig.\,\ref{Setup}). Using the device parameters listed in Table \ref{JM3params} and the transmission matrix model of the device, we plot in Fig.\,\ref{JM3wpt1} (c) and (d), the  calculated amplification curves for the ideal lossless device, which yield a qualitative agreement with the measurement and a bandwidth of $450$ MHz at $10$ dB. 

Saturation power measurements taken in band at the same working point are shown in Fig.\,\ref{JM3wpt1} (e) and (f) for port a and b, respectively. The gain curves plotted versus input power are normalized relative to the pump-off data. As seen in the plots, the coupled-mode JM achieves saturation powers of $-110$ dBm at $7.675$ GHz and $-106$ dBm at $10.745$ GHz, which are comparable to saturation powers attained by some TWPAs \cite{TWPAScience,FloquetTWPA} and are at least $23$ dB higher than typical values measured for $\rm{JM_1}$ (shown in Fig.\,\ref{JM1wpt1}) and resonant-mode JMs implemented using transmission-line resonators \cite{JPCsathighorder}. In Figs.\,\ref{JM3wpt1} (g) and (h), we plot simulation results for the JM device obtained using harmonic-balance in Keysight ADS. In subplot (g), we depict the amplification  response of the JM versus frequency obtained on port a, which, similar to the calculated response, yields a qualitative agreement with the measurement, however with a lower bandwidth of $350$ MHz at $10$ dB. Likewise, the simulated gain versus input power shown in Fig.\,\ref{JM3wpt1} (h) yields a saturation power of $-116$ dBm, which is $6$ dB lower than the measured value. It also features a ``shark-fin" behavior which is absent in the data.

\begin{figure*}
	[tb]
	\begin{center}
		\includegraphics[
		width=2\columnwidth 
		]%
		{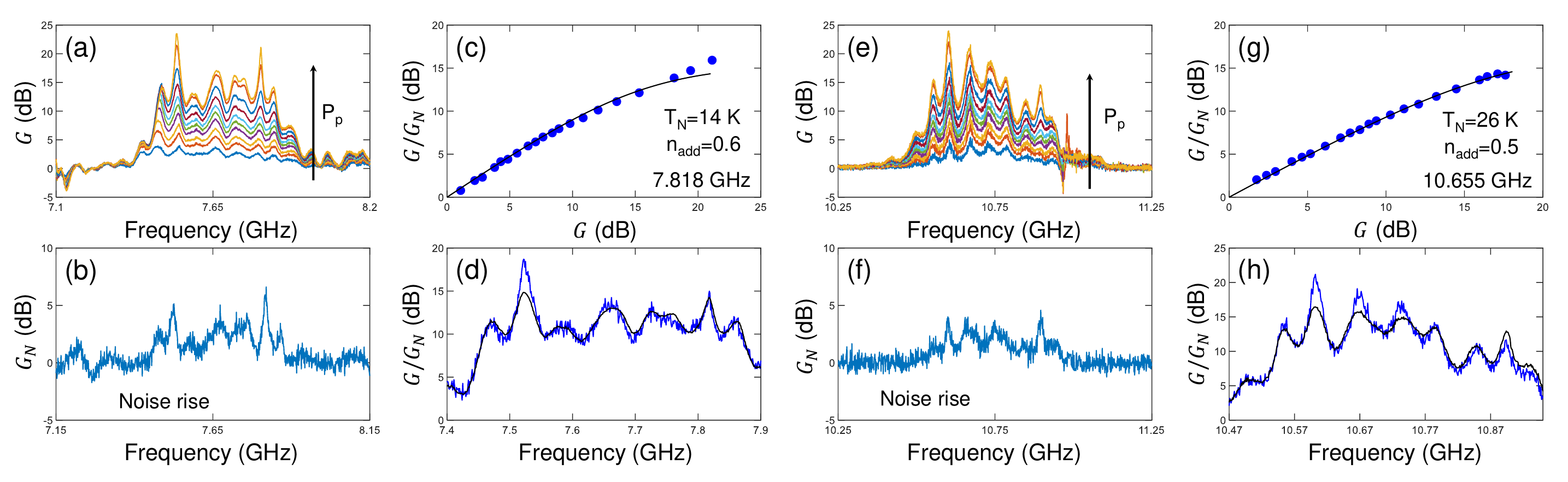}
		\caption{(a) Gain curves measured for mode a of $\rm{JM_3}$ versus frequency for ascending pump power, normalized by the pump off data. (b) Noise rise versus frequency measured for one of the working points shown in (a). (c) SNR improvement versus gain (filled blue circles) measured at $7.818$ GHz. The black curve is a theory fit that employs $T_{\rm{N}}$ and $n_{\rm{add}}$ listed in the plot. (d) SNR improvement for a fixed gain plotted versus frequency (blue). The black curve is a theory fit that uses the same $T_{\rm{N}}$ and $n_{\rm{add}}$ listed in (c). (e)-(h) same as (a)-(d) obtained for mode b of $\rm{JM_3}$          
		}
		\label{JM3QLA}
	\end{center}
\end{figure*}

To further demonstrate that $\rm{JM_3}$ operates near the quantum noise limit, we measure the noise rise $G_N$ of the output chain for port a and b as a function of the device gain $G$. Selected amplification curves measured for different pump powers are shown in Figs.\,\ref{JM3QLA} (a) and (e), which correspond to ports a and b, respectively. An example of $G_N$ versus frequency measured for one of the working points in (a) and (e) is drawn in panels (b) and (f). By plotting the SNR improvement $G/G_N$, versus gain (blue dots) measured at $7.818$ GHz and $10.655$ GHz (shown in panels (c) and (g)), and fitting the data using Eq.\,(\ref{SNR_improv}) (black), we obtain estimates for the effective noise temperature of the output lines $T_N$ (which are consistent with the expected values for the setup) and the added noise by the JM (referred back to the input), i.e., $n_{\rm{add}}$ in the range $0.5-0.6$. Finally, we show in (d) and (h), the SNR improvement versus frequency (blue) plotted for the same working point (pump power) as (b) and (f). The black curves are theory fits, which employ the same extracted parameters displayed in (c) and (g).  
 
\begin{figure*}
	[tb]
	\begin{center}
		\includegraphics[
		width=2\columnwidth 
		]%
		{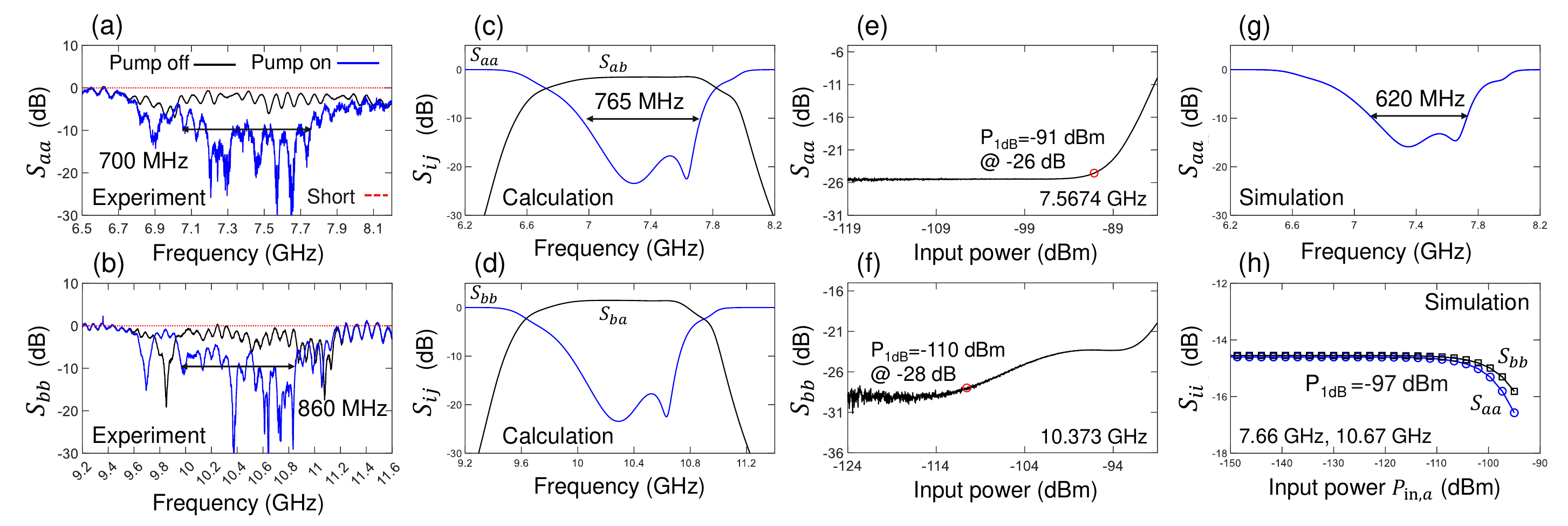}
		\caption{(a) and (b) reflection magnitude $S_{aa}$ and $S_{bb}$ measured versus frequency for $\rm{JM_4}$ with pump on (blue) and pump off (black). Both measured curves are calibrated relative to the reflection magnitude measured for a short ended setup plotted as a $0$ dB guide-to-the-eye dashed red line. The frequency range listed on the plots represent a bandwidth proxy taken at $-10$ dB. (c) and (d) calculated scattering parameters of the device, $S_{aa}$ and $S_{bb}$ in blue and $S_{ab}$ and $S_{ba}$ in black. The different maximum levels of $S_{ba}$ ($1.47$ dB) and $S_{ab}$ ($-1.51$ dB) stem from the frequency conversion process taking place without photon gain. They are also in line with the predictions of the Manely-Rowe relations \cite{ManelyRowe}, i.e., $10\log_{10}\left(\omega_b/\omega_a\right)=1.48$ dB and $10\log_{10}\left(\omega_a/\omega_b\right)=-1.5$ dB, respectively. The theoretical model used in the calculation employs the extracted parameters listed in Table \ref{JM4params}. (e) and (f) saturation power measurements taken at $7.5674$ GHz ($S_{aa}$) and $10.373$ GHz ($S_{bb}$) for the working point of (a) and (b). (g) Keysight ADS simulation of the reflection magnitude $S_{aa}$ versus frequency. (h) Keysight ADS simulation of $S_{aa}$ (blue circles) at $7.66$ GHz and $S_{bb}$ (black squares) at $10.67$ GHz versus the signal power applied to port a.          
		}
		\label{JM4wpt1}
	\end{center}
\end{figure*}

\begin{figure}
	[tb]
	\begin{center}
		\includegraphics[
		width=\columnwidth 
		]%
		{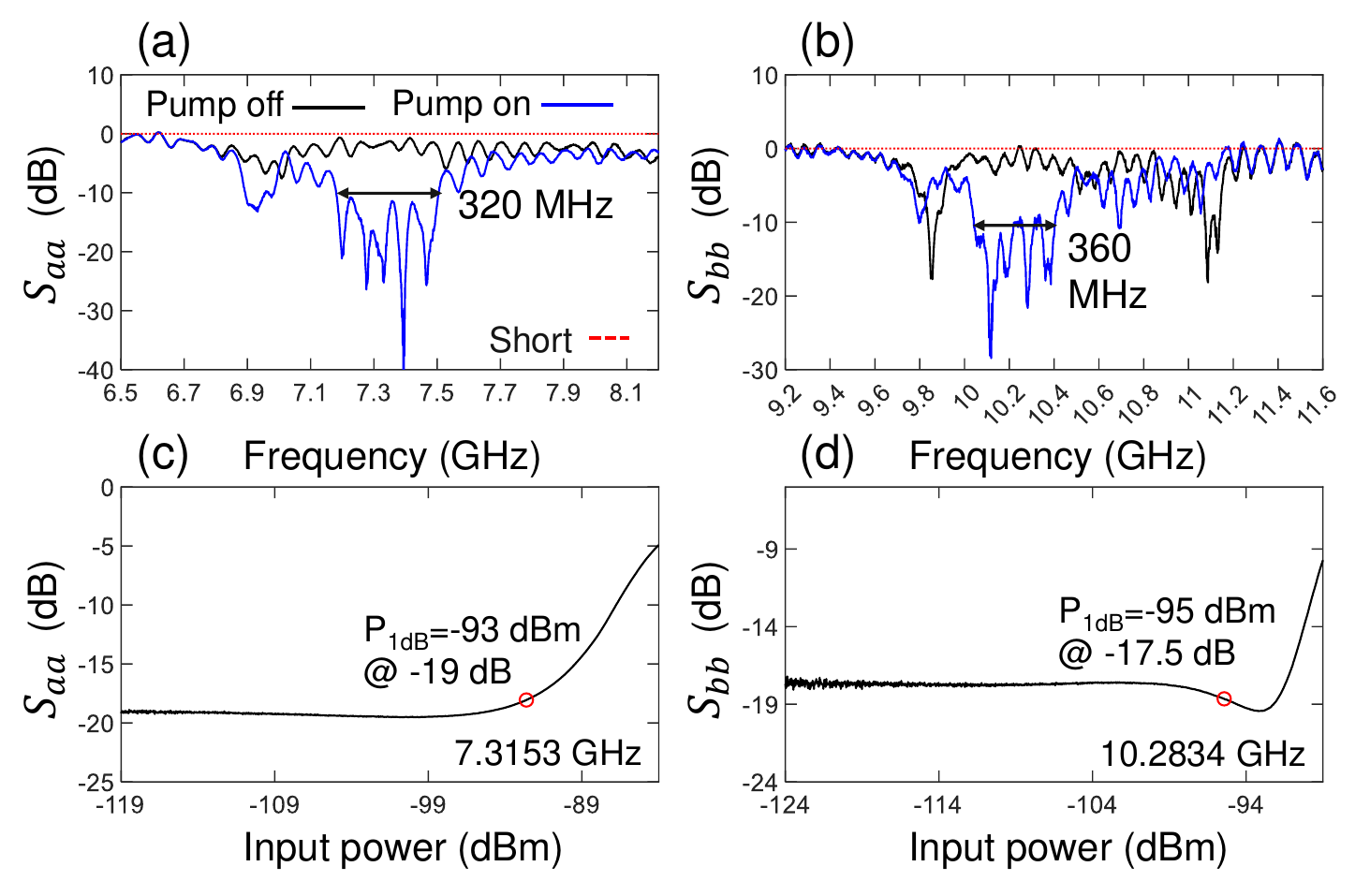}
		\caption{Another measured working point of $\rm{JM_4}$. (a) and (b) show the reflection magnitude $S_{aa}$ and $S_{bb}$ measured versus frequency for $\rm{JM_4}$ with pump on (blue) and pump off (black). Both measured curves are calibrated relative to the reflection magnitude measured for a short ended setup plotted as a $0$ dB guide-to-the-eye dashed red line. The frequency range listed on the plots represent a bandwidth proxy taken at $-10$ dB. (c) and (d) saturation power measurements taken at $7.3153$ GHz ($S_{aa}$) and $10.2834$ GHz ($S_{bb}$) for the working point of (a) and (b).        
		}
		\label{JM4wpt2}
	\end{center}
\end{figure}

Key measurement results obtained using $\rm{JM_4}$ are shown in Figs.\,\ref{JM4wpt1} and Figs.\,\ref{JM4wpt2}. In Fig.\,\ref{JM4wpt1} (a) and (b) we exhibit the amplitude response of $S_{aa}$ and $S_{bb}$ versus frequency measured with pump on (blue) and off (black). Both measurements are normalized relative to the calibrated reflection data measured in-situ for a short termination (red). Relatively wide bandwidth of about $700$ MHz and $860$ MHz is observed at $-10$ dB for the pump on measurement in panels (a) and (b), respectively. A comparable wide bandwidth of about $765$ GHz at $-10$ dB is seen in the calculated response of the device (blue) displayed in Fig.\,\ref{JM4wpt1} (c) and (d). Similar to the calculated response of $\rm{JM_2}$ shown in Fig.\,\ref{JM2wpt1} (i) and (j), we calculate the scattering parameters $S_{ij}$ of $\rm{JM_4}$ using the transmission matrix model outlined in Figs.\,\ref{JRMinverter} and \ref{CircuitModel}, and the device parameters listed in Table \ref{JM4params}. Saturation power measurements corresponding to the working point of (a) and (b) are depicted in Fig.\,\ref{JM4wpt1} (e) and (f), respectively. Relatively high saturation powers of $-91$ dBm and $-110$ dBm are measured for reflection parameters $S_{aa}=-26$ dB and $S_{bb}=-28$ dB, respectively.

In Figs.\,\ref{JM4wpt1} (g) and (h), we exhibit, for comparison, Keysight ADS simulation results obtained for the circuit model of $\rm{JM_4}$. Similar to the calculated $S_{aa}$ amplitude versus frequency shown in (c), the simulation result in (g) exhibits a qualitative agreement with the experimental data shown in (a) and yields a bandwidth of about $620$ MHz at $-10$ dB, which is comparable to the measured value $700$ MHz. Panel (h) exhibits a simulation result of the saturation power at $7.66$ GHz (blue circles) and $10.67$ GHz (black squares) for the same working point and device parameters shown in (g). While the simulated response does not capture the rise of $S_{aa}$ and $S_{bb}$ with input power seen in panels (e) and (f), it yields a saturation power of $-97$ dBm, which lies close to the typical range of $-95$ dBm to $-90$ dBm measured for the device (as seen in Fig.\,\ref{JM4wpt1} (e), and also in Fig.\,\ref{JM4wpt2}(c), (d)).  

Similarly, in Fig.\,\ref{JM4wpt2} we show measurement results of another working point of $\rm{JM_4}$ taken at a slightly different flux and pump frequency than Fig.\,\ref{JM4wpt1} (see Table \ref{ExpCalcSimWpt2}). While the normalized pump-on reflection parameters $S_{aa}$ (a) and $S_{bb}$ (b) exhibit narrower bandwidths, i.e., $320$ MHz and $360$ MHz, the measured saturation powers $-93$ dBm and $-95$ dBm displayed in (c) and (d) are quite similar to the measured value in Fig.\,\ref{JM4wpt1} (e).

\section{Discussion}

The results outlined above show that nondegenerate JM devices operating near the quantum limit can achieve large bandwidth and high saturation power, both in amplification and transduction.

In particular, we obtain bandwidth enhancement in conversion using two design schemes: (1) realizing low external quality-factor, lumped-element, resonant-mode JM devices that are strongly coupled to the external feedlines of modes a and b, and (2) applying impedance matching between the JRM and the external feedlines, by incorporating lumped-element coupled-mode networks. Using the first method, we demonstrate tunable bandwidth of about $600$ MHz for both modes and dynamical bandwidths in the range $220-740$ MHz with calibrated reflection parameters below $-10$ dB. Using the latter scheme, we achieve at the optimal point, dynamical bandwidths of about $700$ MHz and $860$ MHz on ports a and b corresponding to calibrated reflection parameters below $-10$ dB. Similarly, we achieve bandwidth enhancement in amplification using the coupled-mode method, which yields at the optimal point, dynamical bandwidths of about $410$ MHz and $390$ MHz for mode a and b with calibrated reflection gains above $10$ dB.  

For comparison, state-of-the-art resonant-mode JM devices that employ microstrip resonators achieve dynamical bandwidths of about $3-10$ MHz at $20$ dB of gain in amplification \cite{microstripJPC,Roch}, and about $10-20$ MHz at $-20$ dB of reflection in conversion \cite{SMOC}. Thus, the bandwidth obtained using the couple-mode method amounts to an enhancement by approximately a factor of $40$ in amplification and $30$ in conversion.   

Likewise, we achieve concurrent enhancements in the saturation power using several methods. In the case of the resonant-mode JM design, we (1) flux bias the JRM near the Kerr-nulling working point, and (2) dilute the nonlinearity of JRMs by incorporating outer inductors in series with the JRM, thus reducing the participation ratios, and by shunting the JJs of the JRM with relatively small linear inductors ($L_{in}$), which yield $\beta\cong 3.6$. Similarly, in the case of the coupled-mode JM design, we (1) set the parameters of the $LC$ resonators to match a capacitively-shunted JRM operated near the Kerr-nulling working point, (2) dilute the nonlinearity of the JRM by employing $\beta\cong 4.4$, and (3) reduce the pump power needed to attain a desired magnitude of reflection parameters. The latter is a direct result of adding linear $LC$ resonators to the device (i.e., by increasing the order of the ``filter" employed in the implementation) \cite{SynParamCouplNet}.

By utilizing these techniques, we measure in the amplification case of the resonant-mode JM, maximum saturation powers of about $-118$ dBm at $20$ dB of gain. While in the conversion case, we measure as expected, higher saturation powers in the range of $-100$ dBm to $-86$ dBm at a reflection of about $-17$ dB. 

Further enhancement in the saturation power in amplification, by about $10$ dB, is obtained in the coupled-mode JM design, which achieves about $-110$ dBm and $-106$ dBm for port a and b at gains of about $15$ dB and $17$ dB, respectively. When compared to typical saturation powers in the range $-135$ dBm to $-125$ dBm achieved in microstrip-based JM devices \cite{microstripJPC,JPCsathighorder}, the results of the coupled-mode JM amounts to about $15-29$ dB enhancement. Also, notably the achieved saturation power in the coupled-mode design of about $-106$ dBm is on par with saturation powers measured in typical TWPAs in the range $-110$ dBm to $-100$ dBm  \cite{TWPAScience,FloquetTWPA}, while using only $4$ JJs with modest critical currents $2.5$ $\mu$A, as opposed to, for example, a chain of about $2000$ JJs with $4.6$ $\mu$A each \cite{TWPAScience}. 

Similarly, in conversion, we achieve using the coupled-mode JM design high saturation powers in the range $-95$ dBm to $-91$ dBm, measured at a reflection of about $-18$ dB to $-26$ dB.

With such enhancements in the bandwidths and saturation powers, it is interesting to estimate the number of readout signals that can be multiplexed in amplification and conversion using the coupled-mode JMs. Assuming an average frequency spacing between the readout signals of about $70$ MHz, set to ensure an acceptable level of crosstalk in the quantum chip, we obtain in the amplification (conversion) case about $6$ ($11$) signals. Whereas, if we only consider the attainable saturation power while ignoring the bandwidth limitation, and assuming low-power readout, the number of signals which the JMs can potentially process is considerably higher, on the order of tens in amplification and hundreds in conversion.   

Among the disadvantages of lumped-element coupled-mode JMs are: (1) they have numerous design parameters that need to be accurately set within a few percent to obtain a near ideal response. For example, in the $\rm{JM_3}$ and $\rm{JM_4}$ designs, there are about $22$ capacitance and inductance parameters. This makes the response of these devices sensitive to systematic errors in the design (such as underestimating or overestimating the values of the lumped-inductances and capacitances), and sensitive to random fabrication variations of these parameters, especially the capacitances. Moreover, these parameters are not independent, therefore a variation in one can have an unexpected effect on the rest of the circuit. (2) Their response get degraded compared to the ideal lumped-element circuit design due to unintentional parasitic inductances and capacitances forming inside or between sections due to layout limitations. (3) Any mismatch in their lumped-element capacitors and inducators realized along the two arms of $\rm{M_k}$, i.e., $\rm{J_1}$-$\rm{K_1}$ and $\rm{J_2}$-$\rm{K_2}$ (see Fig.\,\ref{Device} (c)), breaks the device symmetry and negatively impacts the differential excitation applied to port $k$.       

It is feasible, however, to overcome and minimize the effect of the first two disadvantages by applying multiple iterations of design, fabrication, and measurement, and by further tightening the variability of the fabrication process. Similarly, it is possible to significantly mitigate the third drawback by combining the respective pairs of capacitive and inductive elements along the two arms of the impedance-matching units $\rm{M_k}$, and differentially excite these units using on-chip $180$ degree hybrids.   

Moreover, since the insertion loss in our devices varying in the range $0-5$ dB is dominated by dielectric loss, this figure can be improved by reducing the dielectric loss of the insulating layers employed in the plate capacitors or replacing them with higher Q materials.   

Finally, it is worth noting that the dynamical bandwidth of $\rm{JM_2}$ can be easily enhanced further by reducing the imbalance in the linear bandwidths of mode a and b, i.e., $347$ MHz, and $912$ MHz, respectively, which stems from the difference between $C_a$ and $C_b$. One straightforward method to achieve that is by reducing the impedance of the external feedlines of port a by half (and therefore reducing the external quality factor of mode a by the same amount), which can be implemented for example by incorporating a quarter-wavelength transformer of impedance $35$ Ohm along the arms of the impedance-matching unit connected to port a (Fig.\,\ref{Device}(b)).    

\section{Conclusion and outlook}

We introduced a blueprint for building large bandwidth and high saturation power lumped-element nondegenerate JMs capable of processing microwave signals near the quantum limit. We implemented the new JM designs using a fabrication process that is based on trilayer Nb junctions and plate capacitors, and introduced a novel transmission line design for pumping the device through a separate port. In particular, we fabricated and measured two JM devices operated in amplification and another two operated in frequency conversion. We achieved significant bandwidth enhancement in the amplification and conversion mode of operation in two JM devices by incorporating impedance-matching networks between the capacitively shunted JRMs and the external feedlines. We realized these impedance-matching networks using coupled-mode networks made of capacitively-coupled $LC$ resonators. We also demonstrated bandwidth enhancement in conversion by realizing a low external quality factor lumped-element, resonant-mode JM device. 

Furthermore, we enhanced the saturation power of the JM devices, by diluting the JRM nonlinearity and designing the coupled-mode networks of the JM devices to have multiple modes per port and operate as close as possible to the Kerr-nulling working point of the JRM.   

Such impedance-matched nondegenerate JM devices, which have enhanced bandwidths of hundreds of megahertz and saturation powers in excess of $-100$ dBm could be useful in a wide range of quantum applications, such as (1) performing fast, frequency-multiplexed, high-fidelity qubit readout in large quantum processors \cite{BuildingLogicalQubits,MultiplexReadoutwithJPA,JPCsqueezingReadout}, (2) serving as wideband and bright sources of continuous-variable entanglement in modular quantum computing architectures and quantum networks \cite{Entswap,CryogenicMwLink,GenEntMwRadoverTL,CVreview}, (3) enabling quantum illumination in the microwave domain \cite{JPCillum,Quanillum}, (4) converting quantum signals from one microwave frequency band to another, and (5) forming nonreciprocal low-loss on-chip Josephson devices that are essential for scaling up quantum systems \cite{ActiveIso,SMOC}.     

\section*{ACKNOWLEDGMENTS} 
We are grateful to John Walter, Jake Fonseca, Khanh Dang, and J. R. Rozen for technical support, Michael Beckley, Charles Rettner, Teddie Magbitang, and Bryan Trimm for preliminary device fabrication, and David Lokken-Toyli, Matthew Beck, and Gerald Gibson for helpful discussions.  

B.A. designed and simulated the devices, performed the measurements, and analyzed the data. D.S. designed the layout, S.C. fabricated the devices, J.N. bump bonded the devices, O.J. contributed to the experimental setup, S.S. and T.M. designed and simulated layout components, V.A. designed the printed circuit boards, C.M. simulated earlier designs and supervised the project. B.A. wrote the paper with input from the other authors.    

\noindent The authors declare no competing interests.

\appendix

\section{Mutual inductance of the JRM}

For the JM, the mutual inductance between modes a and b is primarily parametric with no passive component, $M_0=0$. Thus, it can be written as   
  
\begin{align}
	M\left(  t\right)& =\delta M\cos\left(\omega_pt \right)
	\nonumber
	\\
	& =\dfrac{\delta Me^{j\omega_pt}+\delta M^{*}e^{-j\omega_pt}}{2}, \label{M_t}
\end{align}
  
\noindent where $\omega_p$ is the angular frequency of the pump drive and $\delta M$ is the mutual inductance modulation amplitude, which we infer from the inductive energy of the JRM given by \cite{JPCreview}

\begin{equation}
	E_{\rm{JRM}}=\dfrac{1}{2}L_J\left( I^2_1+I^2_2+\dfrac{1}{4}I^2_p\right) -\dfrac{1}{4} \dfrac{L_JI_{\phi}}{I^2_0}I_1I_2I_p, \label{EJRM1}
\end{equation}

\noindent where $I_1$ and $I_2$ are the rf currents of mode a and b respectively, $I_p$ is the rf current of the pump drive, and $I_0$ is the critical current of the JJs.

By casting Eq.\,(\ref{EJRM1}) in the equivalent form 

\begin{equation}
	E_{\rm{JRM}}=\dfrac{1}{2}L_J\left( I^2_1+I^2_2+\dfrac{1}{4}I^2_p\right) -\dfrac{1}{2}M\left( t\right) I_1I_2, \label{EJRM2}
\end{equation}

\noindent we obtain

\begin{equation}
	\delta M=\dfrac{1}{2}\dfrac{L_JI_{\phi}}{I^2_0}\left| I_p\right|, \label{deltaM}
\end{equation}

\noindent where 

\begin{equation}
	I_{\phi}=I_{0}\sin\left(\varphi_J \right), \label{Icirc}
\end{equation}

\noindent is the circulating current in the outer loop of the JRM, $L_J=L_{J0}/\cos\left(\varphi_J \right) $ is the inductance of the JJs, $L_{J0}=\varphi_0/I_0$ is the JJ inductance at zero external flux, $\varphi_0=\Phi_0/2\pi$ is the reduced flux quantum, and $\varphi_J$ is the superconducting phase difference across the JJs which can be approximated as \cite{Roch}

\begin{equation}
	\varphi_J=\dfrac{\varphi_{e}}{4}-\alpha_J\sin \left( \dfrac{\varphi_{e}}{4}\right)+\dfrac{1}{2}\alpha^2_J\sin\left(\dfrac{\varphi_{e}}{2} \right)  , \label{JJphase}
\end{equation}  

\noindent where $\varphi_{e}=\Phi_{e}/\varphi_0$, $\Phi_e$ is the external flux threading the JRM loop, and $\alpha_J=L_s/L_{J0}$. 

Based on the extracted parameters of the different designs listed in Tables \ref{JM12params}, \ref{JM3params}, and \ref{JM4params}, we estimate $\alpha_J\simeq0.25$ in the case of $\rm{JM}_1$ and $\rm{JM}_2$, and $\alpha_J\simeq0.08$ in the case $\rm{JM}_3$ and $\rm{JM}_4$.

In general, choices for $L_{J0}$, $L_{in}$, $L_{out}$ are made based on design considerations such as optimizing the saturation power of the device or its frequency tunability range. Other factors which play a role in setting their values are layout constraints such as physical spacings between components. These constraints also inevitably set the parasitic inductances $L_s$ in the JRM loop.

\section{Resonant-mode JM design}

\subsection{Flux tunability}
     
Based on Fig.\,\ref{Device}(a), the inductance of the JRM reads \cite{Roch}

\begin{align}
L_{\rm{JRM}}&=\left( L_J+L_s\right) \left| \right| \left( 2L_{in}\right) \nonumber \\ 
&=\dfrac{2L_{in}\left( L_J+L_s\right)}{L_J+L_s+2L_{in}}. \label{LJRM}
\end{align} 

The total inductance of the resonant-mode JM is given by $L_a=L_b=2L_{out}+L_{\rm{JRM}}$, where $L_{out}$ is the series inductance connecting the JRM to the shunt capacitors. $L_{out}$ is usually designed to yield a certain participation ratio for the JM given by \cite{OptJPC}

\begin{equation}
	p=\dfrac{L_{\rm{JRM}}}{L_{\rm{JRM}}+2L_{out}}   , \label{p_ratio}
\end{equation} 

\noindent or set by the geometric inductance of the leads connecting the JRM to the shunt capacitors as realized in the layout. 

For the unpumped resonant-mode JM, the angular resonance frequency of mode $k$, $\omega_k=2\pi f_k$, varies with the external applied flux according to  

\begin{equation}
	\omega_{k}\left( \varphi_e\right) =\dfrac{1}{\sqrt{L_k\left( \varphi_e\right) C_k}}. \label{omega_ab}
\end{equation} 

\subsection{Device parameters}

\begin{figure}
	[tb]
	\begin{center}
		\includegraphics[
		width=\columnwidth 
		]%
		{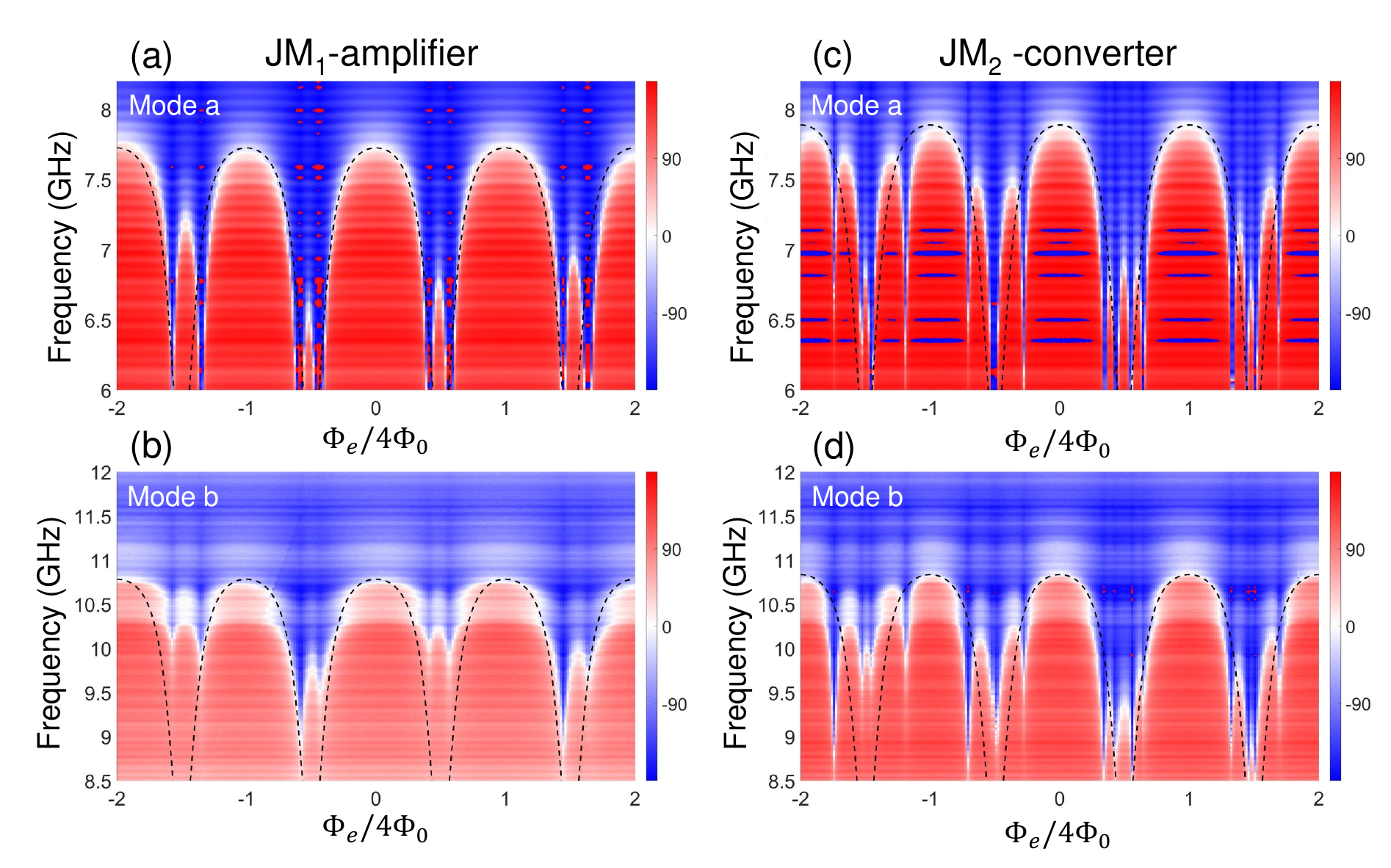}
		\caption{The measured phase of the reflected signals off ports a and b  of $\rm{JM_1}$, shown in (a) and (b) as a function of the normalized external flux. (c) and (d) exhibit similar measurements taken for $\rm{JM_2}$. The resonance frequency of modes a and b follow the zero phase contours. The dashed black curves are theoretical fits to the data obtained using Eq.\,(\ref{omega_ab}).           
		}
		\label{JM12freqvsflux}
	\end{center}
\end{figure}

To extract the lumped-element circuit parameters of $\rm{JM_1}$ and $\rm{JM_2}$ devices (introduced in Fig.\,\ref{Device} (a), (b)), we measure the phase response of mode a and b as a function of applied flux $\Phi_e$ as shown in Fig.\,\ref{JM12freqvsflux}, and apply theory fits to the data based on Eq.\,(\ref{omega_ab}), which are plotted on top of the data using dashed black curves. The circuit parameters used in the fits are listed in Table \ref{JM12params}.

\begin{table}[tbh]
	\centering
	\begin{tabular}{||c |c |c ||} 
		\hline
	Parameter & $\rm{JM}_{1}$ & $\rm{JM}_{2}$  \\
	\hline
	\hline
		$I_0$ ($\mu$A) & $8.1$ & $8.1$     \\ 
		$L_{J0}$ (pH) & $40.6$ & $40.6$    \\
		$L_s$ (pH) & $10$ & $10$    \\
		$L_{in}$ (pH) & $11.3$ & $11.3$  \\
		$L_{out}$ (pH) & $27$ & $27$  \\
		$C_{a}$ (pF) & $6.1$ & $5.85$  \\
		$C_{b}$ (pF) & $3.13$ & $3.13$  \\
		\hline 
	\end{tabular}
	\caption{Circuit parameters of $\rm{JM}_{1}$ and $\rm{JM}_{2}$ devices obtained from the theory fits of the measured frequency response versus applied flux (Fig.\,\ref{JM12freqvsflux}). The variation in $C_a$ between $\rm{JM}_1$ and $\rm{JM}_2$ falls within the fabrication variability range.}
	\label{JM12params}
\end{table}

\begin{figure}
	[tb]
	\begin{center}
		\includegraphics[
		width=\columnwidth 
		]%
		{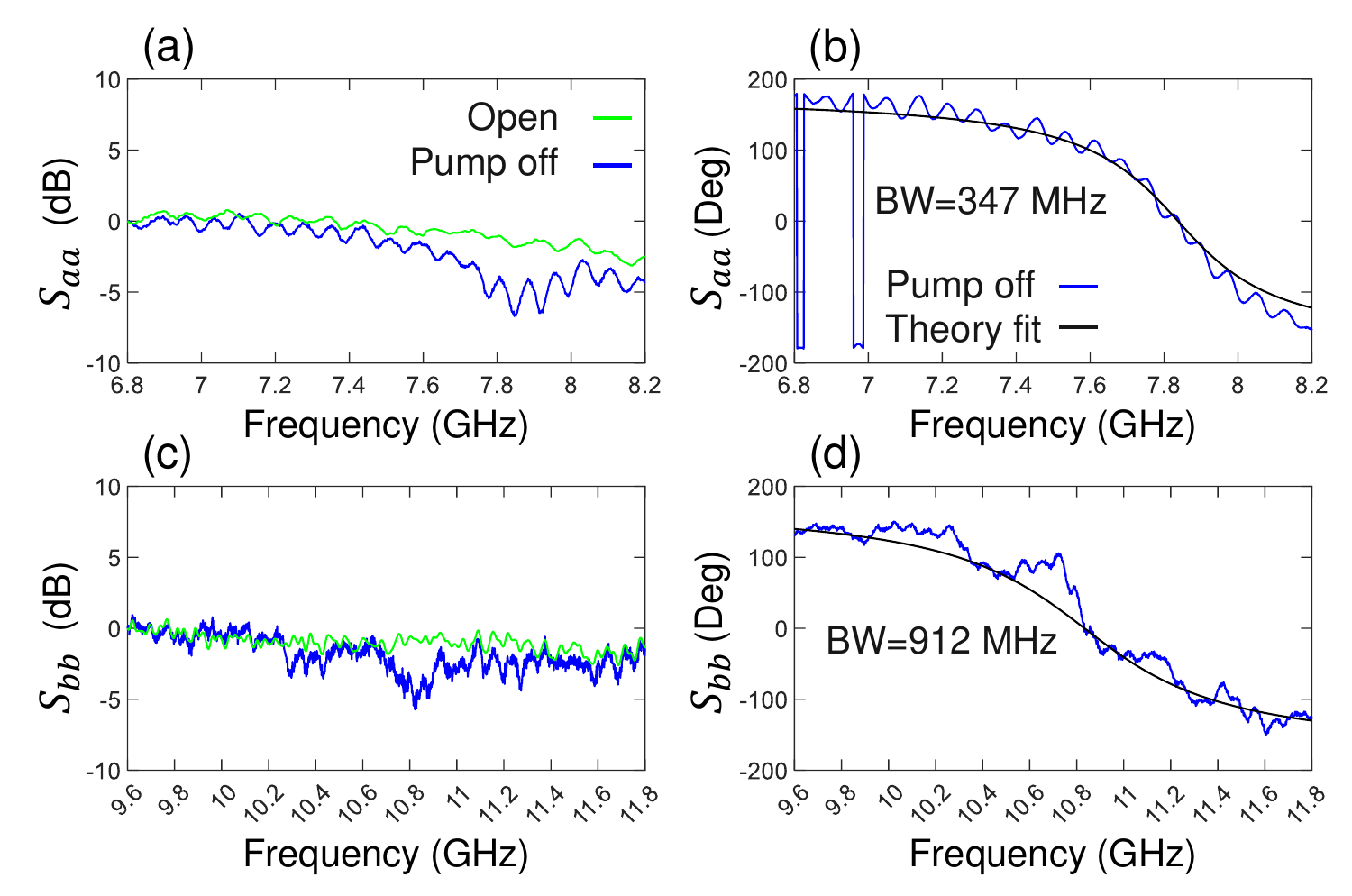}
		\caption{Amplitude (a) and phase (b) of $S_{aa}$ (blue) measured at the maximum resonance frequency of $\rm{JM_2}$. (c) and (d) same as (a) and (b) for $S_{bb}$. The green curves in (a) and (c) are reference levels measured for open ends at nominally identical locations as the mounted device. The black curves in (b) and (d) are theoretical fits based on Eq.\,(\ref{r_a_param_vs_freq}) and Eq.\,(\ref{r_b_param_vs_freq}), which yield linear bandwidths of about $347$ MHz and $912$ MHz for modes a and b, respectively.         
		}
		\label{JM2maxfreq}
	\end{center}
\end{figure}

To obtain estimates for the internal loss and linear bandwidths of $\rm{JM_2}$, we measure the amplitude and phase of modes a and b shown in Fig.\,\ref{JM2maxfreq}, which are taken at the maximum resonance frequency working point. In plot (a) and (c), we exhibit the amplitude of $S_{aa}$ and $S_{bb}$ versus frequency obtained with pump off (blue) and reference curves (green) taken for the same setup and conditions but with open-end terminations (see experimental setup depicted in Fig.\,\ref{Setup}). As seen in the plots, the off-resonance data (pump off) mostly overlaps with the reference level, however, a deviation of $1$ to $4$ dB between the curves, likely due to insertion loss in the device, can be observed on resonance. Likewise, in plot (b) and (d) we exhibit the phase of $S_{aa}$ and $S_{bb}$ versus frequency obtained with pump off (blue). By applying a Lorentzian fit (black) to the phase data, we obtain estimates for the linear bandwidths of mode a and b of about $347$ MHz and $912$ MHz, respectively. 

Similar linear bandwidths of about $400$ MHz and $1$ GHz are measured for mode a and b of $\rm{JM_1}$ (as it shares the same design as $\rm{JM_2}$). 
 
\subsection{JM model}

To obtain the scattering parameters of resonant-mode JM devices such as, $\rm{JM_{1,2}}$, we calculate first their transmission ($ABCD$) matrix, which can be expressed as $\rm{\textbf{T}}_{\rm{total}}=\rm{\textbf{T}_a}\rm{\textbf{T}_{JRM}}\rm{\textbf{T}_b}$. $\rm{\textbf{T}_{JRM}}$ is the transmission matrix of the JRM given in Fig.\,\ref{JRMinverter}(b) for the amplification and conversion modes of operation, whereas $\rm{\textbf{T}_a}$ and $\rm{\textbf{T}_b}$ are the transmission matrices of the parallel $LC$ resonators of mode a (i.e., $L^{\prime}_a$ and $C_a$) and b (i.e., $L^{\prime}_b$ and $C_b$), where        
$L^{\prime}_a=L_a\left( 1-\alpha\right) $, $L^{\prime}_b=L_b\left( 1-\alpha\right) $, and $\alpha\equiv\left| \delta M \right| ^2/4L_aL_b$. 

Both $\rm{\textbf{T}_a}$ and $\rm{\textbf{T}_b}$ can be calculated using the transmission matrix of parallel $LC$ circuits listed in Fig.\,\ref{TransferMatrix}(b).  

As shown in Fig.\,\ref{JRMinverter}(b) the transmission matrix of the JRM in the $J$ shunt representation is set by the parameters

\begin{equation}
	J^{\prime}_{i}=\dfrac{1}{\omega_i\sqrt{L^{\prime}_a}L^{\prime}_b}   \sqrt{\alpha}e^{j\phi_p}, \label{Ji}
\end{equation} 

\noindent where $\phi_p$ is the pump phase and $i=1,2$. It is worth noting that when the pump is off ($\alpha=0$) $L^{\prime}_k=L_k$.

After calculating $\rm{\textbf{T}}_{\rm{total}}$ for given pump parameters, applied flux, and signal frequency, we obtain the scattering parameters of the JM by substituting the $ABCD$ elements of $\rm{\textbf{T}}_{\rm{total}}$ into the following known relations \cite{Pozar}

\begin{align}
	S_{aa}&=\dfrac{A+B/Z_0-CZ_0-D}{A+B/Z_0+CZ_0+D},   \label{S11}
	\\
	S_{ab}&=\dfrac{2\left( AD-BC\right) }{A+B/Z_0+CZ_0+D},   \label{S12}
	\\	
	S_{ba}&=\dfrac{2}{A+B/Z_0+CZ_0+D},   \label{S21}
	\\	
	S_{bb}&=\dfrac{-A+B/Z_0-CZ_0+D}{A+B/Z_0+CZ_0+D}.   \label{S22}
\end{align}   

\section{Coupled-mode JM design}

\subsection{JM model}

Extending the instantaneous (dynamical) bandwidth of Josephson-based amplifiers beyond the amplitude-gain-bandwidth product by incorporating impedance matching elements between the active Josephson-circuit and the external feedlines has been generally successful \cite{JPAimpedanceEng,StrongEnvCoupling}.   

However, applying similar techniques to extend the dynamical bandwidth of JMs is quite challenging for two reasons. First, unlike the other Josephson parametric devices \cite{JPAimpedanceEng,StrongEnvCoupling,BroadbandCPWJPA,BroadbandSnakes,ImpedanceMatchingSNAILs}, which are degenerate, i.e., have a single differential mode and a single physical port, the JM, in contrast, is a nondegnerate device that has two distinct differential modes and two separate physical ports. Second, unlike the other JJ-based or SQUID-based devices, which have two nodes, one of which can be shorted to ground, the JRM at the center of the JM has four main nodes, none of which can be grounded without nullifying its three-wave mixing functionality.  

\begin{table}[tbh]
	\centering
	\begin{tabular}{||c || c | c | c | c | c | c||} 
		\hline
		JM design & $g_{0}$ & $g_{1}$ & $g_{2}$ & $g_{3}$ & $g_{4}$ & $g_{5}$  \\
		\hline
		\hline
		$\rm{JM}_{3}$ & $1$ & $0.76$ & $1.031$ & $1.09$ & $0.34$ & $1.1055$     \\ 
		$\rm{JM}_{4}$ & $1$ & $0.78$ & $1.1$ & $0.75$ & $0.45$ & $0.93$     \\
		\hline 
	\end{tabular}
	\caption{Prototype coefficients used in the design of $\rm{JM}_{3}$ and $\rm{JM}_{4}$. The prototype values are a variation on the Chebyshev coefficients of Eq.\,(\ref{g_arr_thy}), set by optimizing the simulated response of the designs in ADS at $\Phi_e=0.6\Phi_0$.}
	\label{g_params_JM12}
\end{table}

Owing to the symmetrical structure of the JRM, we design matching networks that are mirrored across the symmetry axes of its differential modes a and b as shown in Fig.\,\ref{CircuitModel}. Observe that for the differential mode a (b), the symmetry axis passes through a virtual ground of the shunt capacitor $C_a$ ($C_b$) and the JRM. This yields $L_a=L_b=L_{out}+L_{\rm{JRM}}/2$ for each circuit along the symmetry axis. Furthermore, to make the symmetry and virtual ground aspects more explicit, we substitute in Fig.\,\ref{CircuitModel} the single shunt capacitors implemented across the JRM, i.e., $C_a$ and $C_b$, with their equivalents in the form of two fictitious capacitors in series having double the capacitance, i.e., $2C_a$ and $2C_b$, respectively. Also, in this case $L_J$ in Eq.\,(\ref{deltaM}) is replaced by $L_J/2$.   

Note that such modifications turn each half of the circuit (terminated by the virtual ground) into the standard circuit form discussed in Ref. \cite{SynParamCouplNet}, which can be used to design matching networks that bridge between a parametrically-modulated nonlinear inductor and the respective external ports.

In this work as shown in Fig.\,\ref{CircuitModel}, we design the impedance-matched JMs using the coupled-mode method and implement them using lumped-element parallel \textit{LC} resonators that are coupled by admittance inverters. In particular, we apply a 4-pole Chebyshev prototype for the amplifier and converter designs, whose $i^{\rm{th}}$ coefficient is denoted $g_i$, with the first $g_0$ corresponding to the pumped JRM and the last $g_5$ to the external load. The value of these coefficients, tailored for a certain response such as gain and ripple, can be calculated or found in numerous microwave-design references \cite{SynParamCouplNet,StudyofbroadbandPA,PrototypeBroadPA}. 

For a given frequency bandwidth $\Delta \omega_{k}/2\pi$ of mode $k\in\left\lbrace \rm{a},\rm{b} \right\rbrace $, the corresponding fractional bandwidth is given by $w_{k}=\Delta \omega_{k}/\omega_{k}$, where $\omega_{k}$ is the angular resonance frequency of mode $k$. The values $J_{ij,k}$ of the admittance inverters are calculated using 

\begin{equation}
	J_{ij,k}=\dfrac{w_k}{\sqrt{g_ig_jZ_{i,k}Z_{j,k}}}, \label{Jiip1}
\end{equation} 

\noindent where $i=1,2,3$, $j=i+1$, $Z_{i,k}$ ($Z_{j,k}$) is the impedance of the $i^{\rm{th}}$ ($j^{\rm{th}}$) \textit{LC} resonator. 

For the last admittance inverter preceding the  impedance of the external feedline, i.e., $Z_{5,k}=Z_0$, we have   

\begin{equation}
	J_{45,k}=\sqrt{\dfrac{w_k}{g_4g_5Z_{4,k}Z_{5,k}}}. \label{J45}
\end{equation} 

The capacitors of the admittance inverters are given by 

\begin{equation}
	C_{ij,k}=\dfrac{J_{ij,k}}{\omega_k}, \label{Ciip1}
\end{equation} 

\noindent where $\omega_k/2\pi$ corresponds to the center frequency of the filter on each side of the device.

The inductance $L_{i,k}$ of resonator $i>1$ of mode $k$ with impedance $Z_{i,k}$ is calculated using 

\begin{equation}
	L_{i,k}=\dfrac{Z_{i,k}}{\omega_{k}}, \label{Li}
\end{equation} 

\noindent where $Z_{i,k}$ ($i=2,3, 4$) are degrees of freedom of the design, which are mainly constrained by values that can be conveniently fabricated. For simplicity, we set $Z_{2,k}=50$ Ohm and $Z_{3,k}=\sqrt{Z_{2,k}Z_{4,k}}$. Whereas, $Z_{1,k}$ is set by our choices for $L_k$ and $\omega_k$ since $C^{\prime}_{1,k}=1/\left(\omega^2_k L_k\right)$ and $Z_{1,k}=\sqrt{L_k /C^{\prime}_{1,k}}$. Lastly, to eliminate the admittance inverters between the final shunt \textit{LC} resonator and feedline, we set $Z_{4,k}=w_kZ_{5,k}/\left( g_4g_5\right)$. 

The shunt capacitance $C_{i,k}$ of resonator $i\geq1$ of mode $k$ with impedance $Z_{i,k}$ is calculated using the relations

 \begin{align}
 	C_{1,k}&=C^{\prime}_{1,k}-C_{12,k}=2C_{k},  \label{C1_k}
 	\\
 	C_{2,k}&=C^{\prime}_{2,k}-C_{12,k}-C_{23,k},   \label{C2_k}
 	\\
    C_{3,k}&=C^{\prime}_{3,k}-C_{23,k}-C_{34,k},   \label{C3_k}
 	\\	
 	C_{4,k}&=C^{\prime}_{4,k}-C_{34,k},   \label{C4_k}
 \end{align} 
 
 \noindent where the bare capacitances $C^{\prime}_{i,k}$ for the \textit{LC} sections $i=2,3,4$ read

\begin{equation}
C^{\prime}_{i,k}=\dfrac{1}{\omega_{k}Z_{i,k}}. \label{Cprimei}
\end{equation}
 
For our coupled-mode JM amplifier (converter) design, we use a $4$-pole Chebyshev prototype with $20$ dB ($-20$ dB) of gain (reflection) and $1$ dB of ripple, whose initial values are found in Ref. \cite{SynParamCouplNet}

\begin{align}
	\left\lbrace g_0,..., g_5\right\rbrace 
	=\left\lbrace 1.0, 0.7629, 1.031, 1.1032, 0.3999, 1.1055 \right\rbrace. \label{g_arr_thy}
\end{align}

However, we set the final values for $g_i$ by simulating each coupled-mode lumped-element JM circuit using harmonic-balance in Keysight ADS and tweaking their values based on the simulated response. The modified design coefficients for the amplifier and transducer are listed in Table \ref{g_params_JM12}.   

As to the fractional bandwidths $w_k$, we set them using the so-called ``decrement relation" given by \cite{SynParamCouplNet}

\begin{equation}
	Q=\dfrac{|R_k|}{Z_{1,k}}=\dfrac{g_0g_1}{w_k}, \label{w_R}
\end{equation} 

\noindent where $|R_k|$ is an effective real resistance of the active component (i.e., JRM) at the desired working point and $Q$ is the quality factor of the resonated load, which corresponds to the parallel lumped \textit{LC} resonator with impedance $Z_{1,k}$ shunting the resistor. Using Eq.\,(\ref{w_R}), we obtain   

\begin{equation}
	w_k=g_0g_1\dfrac{Z_{1,k}}{|R_k|}. \label{w_k}
\end{equation} 
  
Similar to the coefficients $g_i$, we optimize the value of $|R_k|$ based on the microwave simulation results of the device. In both the amplification and conversion cases, we set $|R_a|=15$ Ohm and $|R_b|=30$ Ohm.   
  
To derive the scattering parameters of the device, we first calculate the total cascaded \textit{ABCD} matrix of the whole device, which is obtained by multiplying the transmission \textit{ABCD} (\textbf{T}) matrices of the various sections of the circuit denoted $\rm{\textbf{T}}_i$ as shown in Fig.\,\ref{CircuitModel}  

\begin{equation}
	\rm{\textbf{T}}_{\rm{total}}=\prod^{15}_{i=1}\rm{\textbf{T}}_i, \label{ABCDtotal}
\end{equation}

\noindent where the matrices $\rm{\textbf{T}}_i$ ($i\neq8$) representing the coupling capacitors and \textit{LC} resonators are given in Fig.\,\ref{TransferMatrix}, while $\rm{\textbf{T}}_8=T_{JRM}$ representing the JRM is given in Fig.\,\ref{JRMinverter}(b) for the amplifier and transducer cases.  

Afterwards, we substitute the \textit{ABCD} elements of $\rm{\textbf{T}}_{\rm{total}}$ into Eqs.\,(\ref{S11})-(\ref{S22}) to get the scattering parameters of the coupled-mode JM device. 

\subsection{Flux tunability and device parameters}

\begin{figure}
	[tb]
	\begin{center}
		\includegraphics[
		width=\columnwidth 
		]%
		{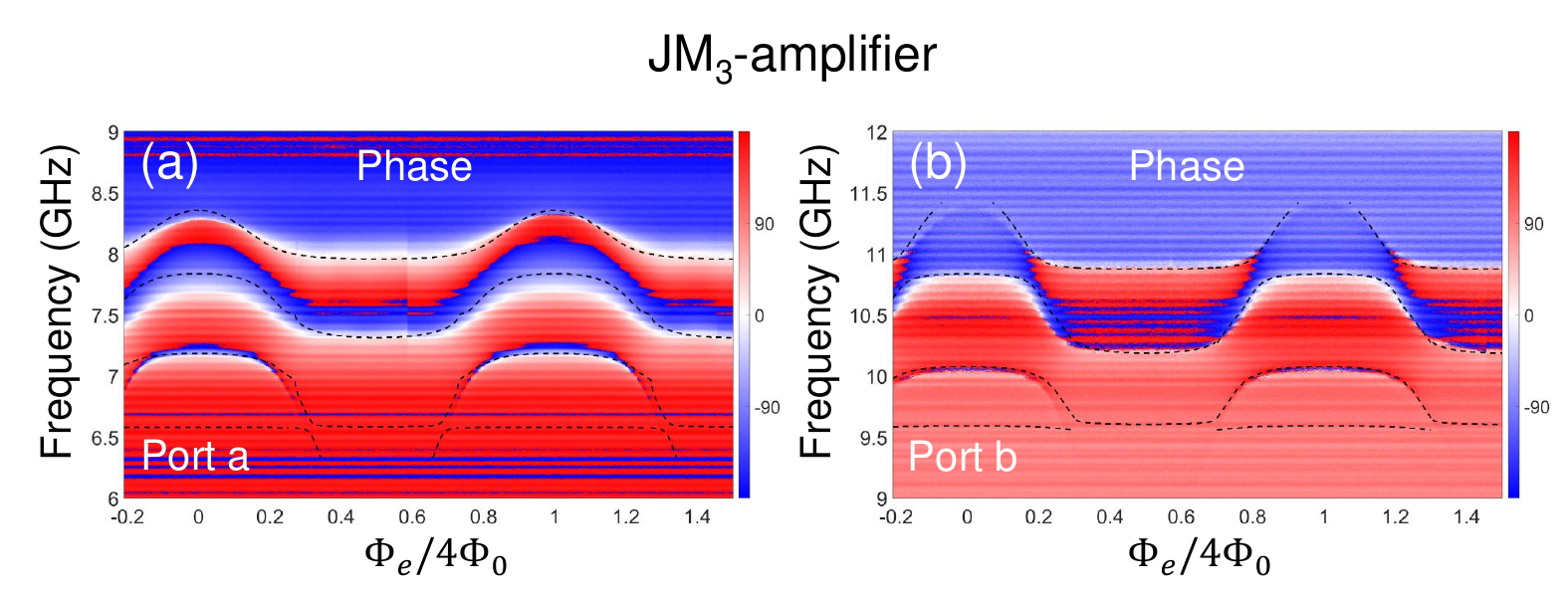}
		\caption{(a) Phase of the reflected signal off port a of $\rm{JM_3}$ measured as a function of frequency and normalized applied flux (Pump off). (b) same as (a) measured for port b. The dashed black curves are theoretical fits. The device parameters used in producing the fits are listed in Table \ref{JM3params}.       
		}
		\label{JM3freqvsflux}
	\end{center}
\end{figure}

Figure\,\ref{JM3freqvsflux} (a) and (b) exhibit the reflection parameter phase measured for port a and b of $\rm{JM_3}$ as a function of frequency and applied external flux. We fit the resonance frequencies of the coupled modes of the device using the theoretical model outlined in Appendix C.1. The calculated fits are plotted on top of the data using dashed black curves. The circuit parameters employed in the fits are listed in Table \ref{JM3params}. Note that since the flux tunability data is taken with no pump $I_p=0$, substituting $\alpha=0$ in the JM model of Appendix C.1 leads to divergence in $\rm{\textbf{T}_{JRM}}$ (i.e., $\rm{\textbf{T}}_8$). To circumvent that, we assume a very small non-zero pump in the calculation (e.g., $\alpha=0.001$).   

\begin{table*}[tbh]
	\centering
	\begin{tabular}{||c c| c c| c c| c c| c c||} 
		\hline
		\multicolumn{10}{|c|}{$\rm{JM}_{3}$ circuit parameters} \\
		\hline
		\hline
		$I_0$ & $2.5$ $\mu$A & $L_{2,a}$ & $1.17$ nH & $C_{2,a}$ & $0.13$ pF & $L_{2,b}$ & $0.8$ nH & $C_{2,b}$ & $0.177$ pF  \\ 
		$L_{J0}$ & $132$ pH & $L_{3,a}$ & $0.597$ nH & $C_{3,a}$ & $0.573$ pF &  $L_{3,b}$ & $0.333$ nH & $C_{3,b}$ & $0.575$ pF \\
		$L_s$ & $10$ pH & $L_{4,a}$ & $ 0.302$ nH & $C_{4,a}$ &  $1.585$ pF & $L_{4,b}$ & $0.132$ nH & $C_{4,b}$ & $1.751$ pF  \\
		$L_{in}$ & $30$ pH &  &  &  $C_{12,a}$ & $0.215$ pF &  &  & $C_{12,b}$ & $0.101$ pF \\
		$L_{out}$ & $9$ pH & & & $C_{23,a}$ & $0.052$ pF & & & $C_{23,b}$ & $0.03$ pF \\
		$C_a$ & $6.276$ pF &  & & $C_{34,a}$ & $0.171$ pF & & & $C_{34,b}$ & $0.124$ pF   \\
		$C_b$ & $3.217$ pF &  & &  &  & & &  &   \\
		\hline 
	\end{tabular}
	\caption{Circuit parameters of $\rm{JM}_{3}$ device obtained from the theory fits of the measured frequency response versus applied flux (Fig.\,\ref{JM3freqvsflux}).}
	\label{JM3params}
\end{table*}

\begin{table*}[tbh]
	\centering
	\begin{tabular}{||c c| c c| c c| c c| c c||} 
		\hline
		\multicolumn{10}{|c|}{$\rm{JM}_{4}$ circuit parameters} \\
		\hline
		\hline
		$I_0$ & $2.5$ $\mu$A & $L_{2,a}$ & $1.131$ nH & $C_{2,a}$ & $0.12$ pF & $L_{2,b}$ & $0.804$ nH & $C_{2,b}$ & $0.171$ pF  \\ 
		$L_{J0}$ & $132$ pH & $L_{3,a}$ & $0.575$ nH & $C_{3,a}$ & $0.582$ pF &  $L_{3,b}$ & $0.343$ nH & $C_{3,b}$ & $0.561$ pF \\
		$L_s$ & $10$ pH & $L_{4,a}$ & $0.292$ nH & $C_{4,a}$ &  $1.517$ pF & $L_{4,b}$ & $0.146$ nH & $C_{4,b}$ & $1.616$ pF  \\
		$L_{in}$ & $30$ pH &  &  &  $C_{12,a}$ & $0.239$ pF &  &  & $C_{12,b}$ & $0.106$ pF \\
		$L_{out}$ & $9$ pH & & & $C_{23,a}$ & $0.073$ pF & & & $C_{23,b}$ & $0.04$ pF \\
		 $C_a$ & $6.287$ pF &  & & $C_{34,a}$ & $0.219$ pF & & & $C_{34,b}$ & $0.142$ pF   \\
		 $C_b$ & $3.219$ pF &  & &  &  & & &  &   \\
		\hline 
	\end{tabular}
	\caption{Circuit parameters of $\rm{JM}_{4}$ device obtained from the theory fits of the measured frequency response versus applied flux (Fig.\,\ref{JM4freqvsflux}).}
	\label{JM4params}
\end{table*}

\begin{figure}
	[tb]
	\begin{center}
		\includegraphics[
		width=\columnwidth 
		]%
		{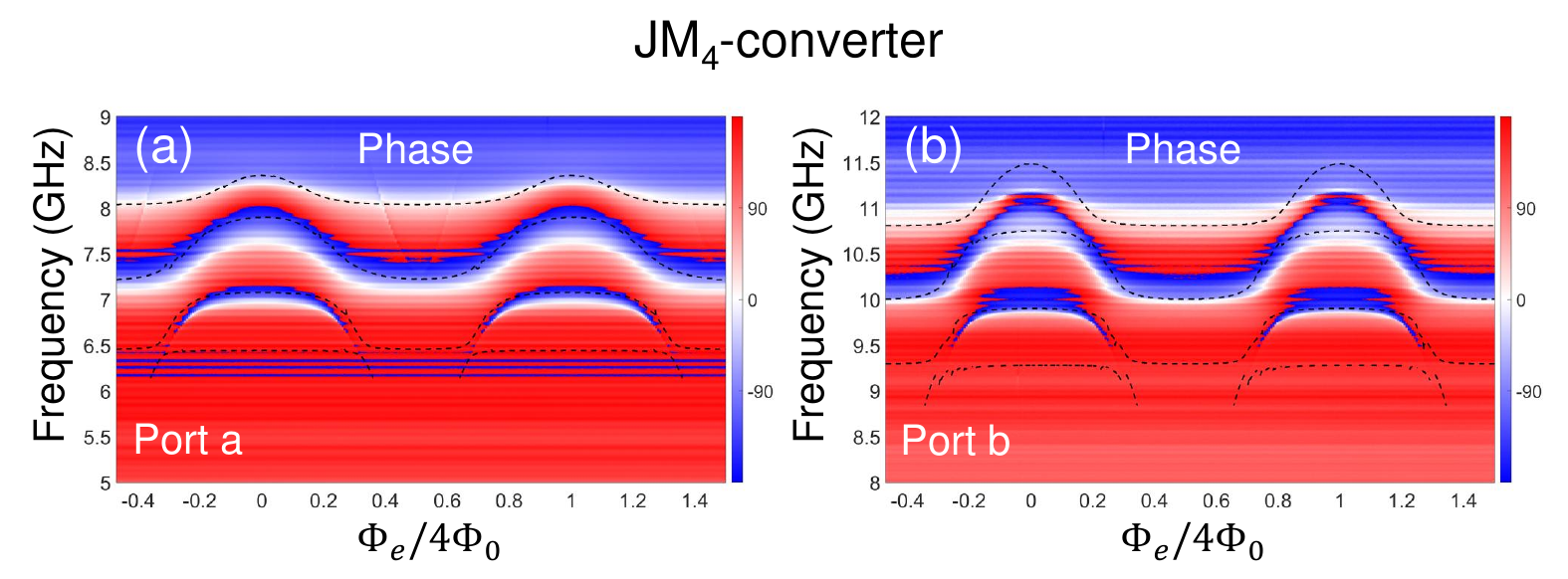}
		\caption{(a) Phase of the reflected signal off port a of $\rm{JM_4}$ measured as a function of frequency and normalized applied flux (Pump off). (b) same as (a) measured for port b. The dashed black curves are theoretical fits. The device parameters used in producing the fits are listed in Table \ref{JM4params}.         
		}
		\label{JM4freqvsflux}
	\end{center}
\end{figure}

Figure\,\ref{JM4freqvsflux} (a) and (b) exhibit the reflection parameter phase measured for port a and b of $\rm{JM_4}$ as a function of frequency and applied external flux. Similarly, we fit the resonance frequencies of the coupled modes of the device using the theoretical model outlined in Appendix C.1. The circuit parameters employed in the fits are listed in Table \ref{JM4params}.

It is important to point out that due to the large number of design parameters associated with $\rm{JM}_{3,4}$, we impose three restrictions in the fitting process, (1) we set the parameter values associated with the JRM and $L_{out}$ to be identical (based on the best fits applicable to both devices), (2) we assume that for each port (i.e., A and B) the lumped-element matching networks realized along the arm $\rm{J}_1$ and $\rm{K}_1$ are identical to those along $\rm{J}_2$ and $\rm{K}_2$ (as defined in Fig.\,\ref{Device} (c)), and (3) we limit the possible variation of the various parameter values to less than $\pm10\%$ compared to the respective ideal original design. 

\section{JM simulation model}

\begin{table*}[tbh]
	\centering
	\begin{tabular}{|c | c| c c c| c c| c | c c c|} 
		\hline
		& \multicolumn{6}{|c|}{$\rm{JM}_{1}$ results} & \multicolumn{4}{|c|}{$\rm{JM}_{2}$ results} \\
		\hline
		\hline
		Fig. & \ref{JM1wpt1}(c)(d) & \ref{JM1wpt1}(i)(j) & \ref{JM1wpt1}(k)(l) & \ref{JM1wpt1}(p)(q) & \ref{JM1wpt2}(a)(b) & \ref{JM1wpt2}(c)(d) & \ref{JM2wpt1}(c)(d) & \ref{JM2wpt1}(g)(h) & \ref{JM2wpt1}(i)(j) & \ref{JM2wpt1}(m)(n) \\ 
		\hline
		$\Phi_{e}/\Phi_{0}$ & $1.45$ & $1.24$ & $0.7$ & $0.7$ & $1.6$ &  $1.5$ & $1.06$  & $0.93$ & $1.2$ & $0.8$ \\
		$f_p$ (GHz) & $17.381$ & $17.484$ & $17.5$ & $18$ & $16.506$ &  $16.3$ & $2.9$ & $2.996$ & $2.8$ & $2.8$ \\
		\hline
	\end{tabular}
	\caption{External flux and pump frequency applied in the measurement, calculation, and simulation of $\rm{JM_1}$ and $\rm{JM_2}$.}
	\label{ExpCalcSimWpt1}
\end{table*}

\begin{table*}[tbh]
	\centering
	\begin{tabular}{|c | c c c | c c c | c |} 
		\hline
		& \multicolumn{3}{|c|}{$\rm{JM}_{3}$ results} & \multicolumn{4}{|c|}{$\rm{JM}_{4}$ results} \\
		\hline
		\hline
		Fig. & \ref{JM3wpt1}(a)(b) & \ref{JM3wpt1}(c)(d) & \ref{JM3wpt1}(g)(h) & \ref{JM4wpt1}(a)(b) & \ref{JM4wpt1}(c)(d) & \ref{JM4wpt1}(g)(h) & \ref{JM4wpt2}(a)(b) \\ 
		\hline
		$\Phi_{e}/\Phi_{0}$ & $0.776$ & $1$ & $0.75$ & $0.45$ & $0.95$ &  $0.75$ & $0.4$   \\
		$f_p$ (GHz) & $18.418$ & $18.2$ & $18.41$ & $3.12$ & $3$ &  $3.01$ & $2.86$  \\
		\hline
	\end{tabular}
	\caption{External flux and pump frequency applied in the measurement, calculation, and simulation of $\rm{JM_3}$ and $\rm{JM_4}$.}
	\label{ExpCalcSimWpt2}
\end{table*}

As stated in the main text, the simulation results are obtained using the harmonic-balance tool in Keysight ADS (2023 version). The circuit schematic applied in the simulation is identical to the lumped-element circuit displayed in Fig.\,\ref{Device}(a) in combination with the trivial matching network shown in (b) in the case of $\rm{JM_{1}}$ and $\rm{JM_{2}}$ or the coupled-mode matching network exhibited in (c), in the case of $\rm{JM_{3}}$ and $\rm{JM_{4}}$. Also, similar to the experimental setup shown in Fig.\,\ref{Setup}, we excite ports a and b of the JM differentially using ideal $180$ degree hybrids. 

The JJs of the JRM are modeled using the JJ3 circuit element. To mimic the effect of flux biasing in the loop, we apply a circulating current in the JRM by connecting an ideal dc-current source shunted by a large capacitor ($1$ F) in series with each JJ. The current of the dc source is set by the corresponding superconducting current flowing in the outer loop, $I_{\phi}$ given by Eq.\,(\ref{Icirc}), whereas the large shunt capacitor is added to act like an open at dc (across the current source) and a short at microwave frequencies.

It is important to point out that the harmonic-balance simulation we use suffers from convergence limitations, which prevent us from fully simulating the JM circuit at any desired working point. In particular, the simulation breaks down when applying magnetic fluxes beyond $0.75-0.8\Phi_0$, or when the input signal power exceeds the saturation power by a few dBm, or when the applied pump power becomes ``appreciable", depending on the circuit, pump frequency, and applied flux. Thus, due to these convergence limitations, our optimization of the JM design parameters is mainly done by simulating their operation near $0.6\Phi_0$.     

In Tables \ref{ExpCalcSimWpt1}, \ref{ExpCalcSimWpt2}, we list the external fluxes and pump frequencies, which we employ to obtain the various measurement results, calculated responses, and simulated curves exhibited in the main text. The flux and pump parameters employed in the calculation and simulation were generally treated as degrees of freedom, whose values are chosen based on the resultant agreement between the measured and generated response with respect to their frequency range, shape, and magnitude. 

\section{Setup}

\begin{figure*}
	[tb]
	\begin{center}
		\includegraphics[
		width=2\columnwidth 
		]%
		{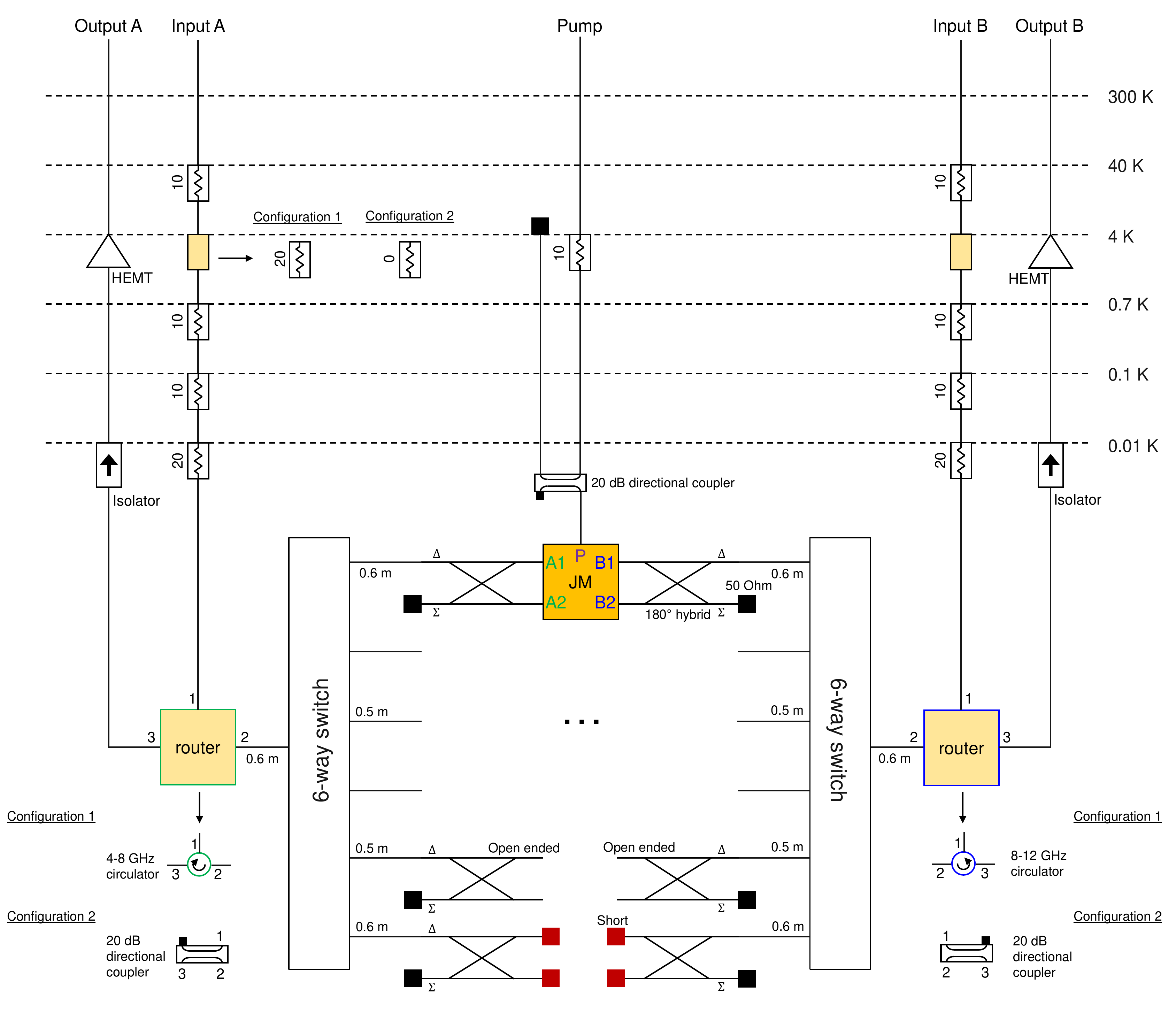}
		\caption{Experimental setup used in the characterization and measurement of the different JMs. Ports a and b of the tested JM are connected to Krytar $180$ degree hybrids used for exciting the differential modes a and b (the input and output signals pass through the delta port). The pump drives are directly fed to the JM devices and carried by separate input lines. Two 6-way switches are used to switch between JMs of different designs and $180$ degree hybrids that are terminated by an open or a short are used for calibration measurements. $\rm{JM_1}$ and $\rm{JM_2}$ are measured using configuration 1, in which the switch assigned to port a (b) of the JMs is connected through a circulator to the input and output lines of mode a (b), and a $20$ dB attenuator is included in the input line A (B) at the $4$ K stage. Whereas, $\rm{JM_3}$ and $\rm{JM_4}$ are measured using configuration 2, in which the switch assigned to port a (b) of the JMs is connected through a 20 dB directional coupler to the input and output lines of mode a (b), and no attenuator is included in the input line A (B) at the $4$ K stage.
		}
		\label{Setup}
	\end{center}
\end{figure*}

A schematic diagram of the fridge setup used in the measurement of the JM devices is shown in Fig.\,\ref{Setup}. It includes two input and output lines for probing ports a and b of the JMs. As seen in the diagram the input and output lines are joined via routers that separate the incoming and outgoing microwave signals reflecting off the JM ports. In one experimental version, the signal router is realized using a three-port magnetic circulator, while in another it is implemented using a $20$-dB directional coupler (whose ``isolated" port is terminated by a $50$ Ohm). We use the former, denoted configuration 1, to measure $\rm{JM_1}$ and $\rm{JM_2}$, and the latter, denoted configuration 2, to measure $\rm{JM_3}$ and $\rm{JM_4}$. Due to these changes in the setup, the two groups of JM devices are characterized in two separate cooldowns. To balance the total attenuation of the input lines, we incorporate $20$ dB attenuators at the $4$ K stage when using the circulators. These added attenuators compensate for the  attenuation of $20$ dB introduced by the directional couplers in configuration 2.  

As seen in the setup diagram, to excite the differential modes a (b) of the JMs, we connect their ports $A_1$ and $A_2$ ($B_1$ and $B_2$) to commercial wideband $180$ degree hybrids (Krytar $6-20$ GHz) and probe their response through the delta ($\Delta$) ports of the couplers, while terminating the sum ($\Sigma$) ports of the hybrids by $50$ Ohm cold loads. 

Furthermore, to enable testing of more than one JM device per cooldown, we use two 6-way Radiall switches, whose common ports connect to the routers assigned for measuring ports a and b of the JMs. In addition to the JM devices under test, we allocate entries on the switches for calibrating the reflection parameters in-situ, which we connect to coax cables and hybrids that are nominally identical to the ones used in the JM measurement, terminated instead by an open or a short. 

Due to necessary spacings and mounting constraints of the various components at the base temperature stage, relatively long coaxial lines $0.5-0.6$ m are used to connect between the different components. Examples of such constraints include housing the JM devices deep inside a cryoperm magnetic shield and mounting the JM devices away from the magnetic components, i.e., isolators and circulators. However, incorporating such relatively long coaxial lines between the intermediate parts at the base stage gives rise to unwanted ripples in the measured frequency response of the JMs, originating from standing waves in the lines due to mismatches in the impedance of the various ports, especially those belonging to the routers and samples (see Appendix G for more details).    

The pump drives are fed to the JMs though separate input lines, which include wideband $20$ dB directional couplers at the base stage. The couplers attenuate the pump tones that drive the JMs by directing a large portion of their powers towards the $4$ K stage where they get dissipated in $50$ Ohm terminations. 

\section{Scattering parameters of resonant-mode JMs operated in conversion} 

For completeness, we present here the reflection and transmission parameters used in producing the theoretical results shown in Fig.\,\ref{ConvBW} and fits in Fig.\,\ref{JM2maxfreq} (b), (d). For resonant-mode JMs operated in conversion, the transmission parameter can be written as \cite{JPCreview}   
 
\begin{align}
	S_{ab}=\dfrac{2i\rho e^{-i\phi_p}}{\chi_{a}^{-1}\chi_{b}^{-1}+ \rho ^2}, \label{Sab_param_vs_freq} \\ 
	S_{ba}=\dfrac{2i\rho e^{i\phi_p}}{\chi_{a}^{-1}\chi_{b}^{-1}+ \rho ^2}, \label{Sba_param_vs_freq}
\end{align}

\noindent where $\rho\geq 0$ is a dimensionless parameter representing the pump amplitude and $\chi's$ are the bare response functions of modes a and b, whose inverses depend linearly on the signal and idler frequencies $f_{\rm{1}}$ and $f_{\rm{2}}$, respectively: 

\begin{align}
	\chi_{a}^{-1}[\omega_{\rm{1}}]=1-2i\dfrac{\omega_{\rm{1}}-\omega_{a}}{\gamma_{a}}, \nonumber \\ 
	\chi_{b}^{-1}[\omega_{\rm{2}}]=1-2i\dfrac{\omega_{\rm{2}}-\omega_{b}}{\gamma_{b}}, 
	\label{Chi_params}%
\end{align}

\noindent where $\omega_{\rm{1}}=2\pi f_{1}$, $\omega_{a}=2\pi f_{a}$ and the applied pump angular frequency is given by $\omega_{p}=\omega_{b}-\omega_{a}=\omega_{\rm{2}}-\omega_{\rm{1}}$, where $\omega_{b}=2\pi f_{b}$ and $\omega_{\rm{2}}=2\pi f_{2}$.

Similarly, the JM reflection parameters at ports `a' and `b', can be written as 

\begin{align}
	S_{aa}[\omega_{\rm{1}}]=\dfrac{\chi_{a}^{-1*}\chi_{b}^{-1}-\rho ^2}{\chi_{a}^{-1}\chi_{b}^{-1}+ \rho ^2}, \label{r_a_param_vs_freq} \\  
	S_{bb}[\omega_{\rm{2}}]=\dfrac{\chi_{a}^{-1}\chi_{b}^{-1*}- \rho ^2}{\chi_{a}^{-1}\chi_{b}^{-1}+ \rho ^2}. 
	\label{r_b_param_vs_freq} 
\end{align}

\section{Effect of interference and multiple reflections on the measured response}

\begin{figure}
	[tb]
	\begin{center}
		\includegraphics[
		width=\columnwidth 
		]%
		{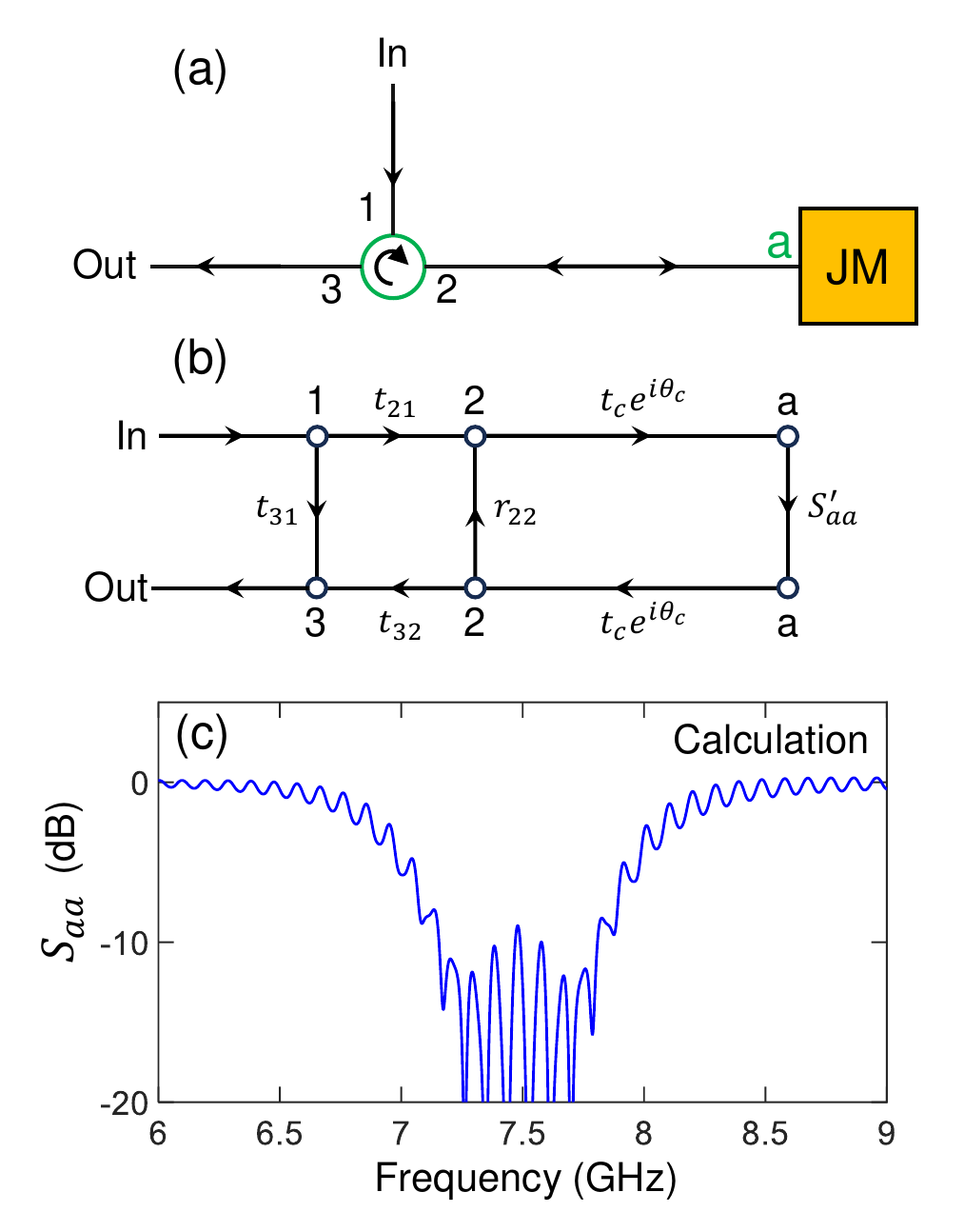}
		\caption{(a) Simplified diagram of the reflection measurement setup. (b) The corresponding signal flow graph. (c) Normalized reflection response calculated for the working point of $\rm{JM}_{2}$ exhibited in Fig.\,\ref{JM2wpt1}(g). In the calculation, we model the effect of reflections and wave interference taking place in the setup based on the signal flow graph shown in (b) (see text for more details). The parameter values employed in the calculation are, $t=t_{21}=t_{32}=t_{c}=0.95$, $t_{31}=0.1$, $r_{22}=0.17$, $l_c=1.1$ m, and $\varepsilon=2.1$.         
		}
		\label{RippleModelEffect}
	\end{center}
\end{figure}

Most ripples seen in the reflection measurements of the JM devices, especially those with large bandwidths, originate from interference effects caused by multiple reflections in the experimental setup; primarily between the JM ports and the corresponding signal ``routers" (i.e., circulators or directional couplers shown in Fig.\,\ref{Setup}), which are connected via coaxial cable lines with lengths of about $1.1-1.2$ m. 

To demonstrate the pronounced ripple effect generated in the reflection measurement setup, we use a simple interference model depicted in Fig.\,\ref{RippleModelEffect}. In panel (a) we sketch a simplified diagram of the reflection measurement setup; consisting of an input line connected to port $1$ of a three-port circulator, a JM device whose port a is connected to port $2$ of the circulator via a relatively long coaxial cable, and an output line connected to port $3$ of the circulator. In panel (b), we draw the corresponding signal flow graph for the simplified reflection setup presented in (a). Next to the arrows indicating the possible paths taken by the signals propagating between the input and output lines, we list the reflection or transmission amplitudes associated with the different path sections. In the graph, $t_{21}$ and $t_{32}$ represent the transmission amplitudes through the circulator from port 1 to 2 and 2 to 3, which are aligned with its circulation direction. In contrast, $t_{31}$ represents the transmission amplitude from port 1 to 3 of the circulator, which corresponds to transmission in the opposite direction. Similarly, $t_c$ and $\theta_c$ represent the transmission amplitude and accumulated phase through the coaxial lines connecting the circulator and the JM port. While $r_{22}$ and $S^{\prime}_{aa}$ represent the amplitude of the reflection off port 2 of the circulator and port a of the JM, respectively.

By inspection, we can write down the effective reflection parameter amplitude $S_{aa}$ measured in the experiment       

\begin{align}
	S_{aa}=t_{31}+\dfrac{t_{32}t_{21}t^2_ce^{2i\theta_c}S^{\prime}_{aa}}{1-r_{22}S^{\prime}_{aa}t^2_{c}e^{2i\theta_c}}, \label{S_aa_ripples}
\end{align}

\noindent where $\theta_c=2\pi f_1\sqrt{\varepsilon}l_c/c$, $c$ is the speed of light in vacuum, $l_c$ is the length of the coaxial cables, and $\varepsilon$ is the relative permittivity constant of the dielectric material in the cables. Although Eq.\,(\ref{S_aa_ripples}) is derived for port a of the JM, a similar relation can be obtained for reflections off port b. 

For simplicity, we assume that all transmission and reflection parameter amplitudes that are not related to the JM, i.e., $t_{31}$, $t_{32}$, $t_{21}$, $t_{c}$, $r_{22}$ are real and independent of frequency. Whereas, $S^{\prime}_{aa}$ is calculated for each input frequency using Eq.\,(\ref{S11}) or Eq.\,(\ref{r_a_param_vs_freq}) depending on the particular case. 

As an example, we apply Eq.\,(\ref{S_aa_ripples}) to the working point of $\rm{JM_2}$ presented in Fig.\,\ref{JM2wpt1}(g). Since the pump-on reflection measurement shown in Fig.\,\ref{JM2wpt1}(g) is normalized by the pump off measurement, we plot in Fig.\,\ref{RippleModelEffect}(c) the calculated normalized response $|S_{aa}|^2=|S^{\rm{on}}_{aa}/S^{\rm{off}}_{aa}|^2$, where both $S^{\rm{on}}_{aa}$ and $S^{\rm{off}}_{aa}$ are calculated using Eq.\,(\ref{S_aa_ripples}) (which correspond to different $S^{\prime}_{aa}$). In the $S^{\rm{off}}_{aa}$ case, we employ Eq.\,(\ref{r_a_param_vs_freq}) to calculate $S^{\prime}_{aa}$ for $\rho=0$, $\gamma_{a}/2\pi=400$ MHz, and $\gamma_{b}/2\pi=900$ MHz, whereas in the $S^{\rm{on}}_{aa}$ case, we employ Eq.\,(\ref{S11}) (whose result is plotted in Fig.\,\ref{JM2wpt1}(i)). In the calculation we assume that the JM is lossless and that the circulator has an insertion loss of about $-0.45$ dB in the forward direction, power reflection of about $-15$ dB, and power isolation of $20$ dB. We also assume that the coaxial cables have a similar insertion loss in each direction $-0.45$ dB. 

As seen in Fig.\,\ref{RippleModelEffect}(c), the ripples in the calculated response can be as significant as the ones observed in the measured data especially in band. We also note that the average ripple separation in frequency is about $90$ MHz in the calculation compared to about $80$ in the experiment. While understandably the simple model presented in this section cannot produce every feature in the data, it generally explains most of the ripples and dips seen in the reflection curves.

\section{Added noise in amplification}

To show that the nondegenerate Josephson amplifiers, i.e., $\rm{JM_1}$ and $\rm{JM_3}$, operate near the quantum limit, we measure the improvement in the signal-to-noise ratio of the output chain of port $k$, which is given by $G/G_{N,k}$, where $G$ is the power gain of the JM and $G_{N,k}$ is the output noise ratio (or noise rise) corresponding to the JM being on versus off

\begin{align}
	G_{N,k}=\dfrac{T_{N,k}+GT_{Q,k}n_{\rm{vac}}+\left( G-1\right)T_{Q,k}n_{\rm{add}} }{T_{N,k}+T_{Q,k}}, \label{G_N}
\end{align}

\noindent where $T_{Q,k}=\hbar \omega_k/ k_B$ is the equivalent temperature of a mode $k$ photon, $T_{N,k}$ is the effective noise temperature of the output chain for mode $k$ in the absence of the JM, $n_{\rm{vac}}=1/2$ is input vacuum noise, and $n_{\rm{add}}$ is the number of noise equivalent input photons added
by the JM.

Using Eq.\,(\ref{G_N}), the signal-to-noise ratio improvement due to the JM reads

\begin{equation}
	\dfrac{G}{G_{N,k}}=\dfrac{T_{N,k}+T_{Q,k}}{T_{N,k}/G+T_{Q,k}\left[n_{\rm{vac}}+n_{\rm{add}}\left(G-1\right)/G\right] }. \label{SNR_improv}
\end{equation}

\end{document}